\documentclass[aps,prx,twocolumn,superscriptaddress,longbibliography]{revtex4-2}
\usepackage{amsmath}
\usepackage{amsfonts}
\usepackage{MnSymbol}
\usepackage{hyperref}
\hypersetup{
	colorlinks=true,
	linkcolor=blue,
	filecolor=blue,
	citecolor = black,      
	urlcolor=cyan,
}

\usepackage{graphicx}
\usepackage{siunitx}
\graphicspath{{./Figures/}}

\usepackage{xargs}                      
\usepackage[pdftex,dvipsnames]{xcolor}
\usepackage{color}
\usepackage{braket}
\usepackage[normalem]{ulem}

\newcommand{\magn}[1]{\left|#1\right|}

\newcommand{\adj}[1]{#1^{\dagger}}

\newcommand{\figref}[1]{Fig.~\ref{#1}}
\newcommand{\appref}[1]{Appendix~\ref{#1}}

\newcommand{\eqeqref}[1]{Eq.~\ref{#1}}

\newcommand{\bbra}[1]{\left \llangle #1 \right \rvert }
\newcommand{\kket}[1]{\left \lvert #1 \right \rrangle }

\newcommand{\ph}{\phi^{a}}
\newcommand{\pv}{\phi^{b}}
\newcommand{\px}{\phi^{c}}

\newcommand{\normfacWrAA}{J_{s,\text{WrAA}}}

\begin{document}


\title{Onset of scrambling as a dynamical transition in tunable-range quantum circuits}
\author{Sridevi Kuriyattil}
\thanks{These two authors contributed equally}
\address{Department of Physics and SUPA, University of Strathclyde, Glasgow G4 0NG}
\author{Tomohiro Hashizume}
\thanks{These two authors contributed equally}
\address{Department of Physics and SUPA, University of Strathclyde, Glasgow G4 0NG}
\author{Gregory Bentsen}
\address{Martin A. Fisher School of Physics, Brandeis University, Waltham, Massachusetts 02465, USA}
\author{Andrew J. Daley}
\address{Department of Physics and SUPA, University of Strathclyde, Glasgow G4 0NG}
\begin{abstract}
In a fast scrambling many-body quantum system, information is spread and entanglement is built up on a timescale that grows logarithmically with the system size. This is of fundamental interest in understanding the dynamics of many-body systems, as well as in efficiently producing entangled resource states and error-correcting codes. In this work, we identify a dynamical transition marking the onset of scrambling in quantum circuits with different levels of long-range connectivity. In particular, we show that as a function of the interaction range for circuits of different structures, the tripartite mutual information exhibits a scaling collapse around a critical point between two clearly defined regimes of different dynamical behaviour. We study this transition analytically in a related long-range Brownian circuit model and show how the transition can be mapped onto the statistical mechanics of a long-range Ising model in a particular region of parameter space. This mapping predicts mean-field critical exponents $\nu = -1/(1+s_c)$, which are consistent with the critical exponents extracted from Clifford circuit numerics. In addition to systems with conventional power-law interactions, we identify the same phenomenon in deterministic, sparse circuits that can be realised in experiments with neutral atom arrays.    
\end{abstract}

\maketitle




\section{Introduction}
The question of how fast information is spread in a quantum system and how quickly entanglement is generated is relevant not only in understanding many-body dynamics, but especially in noisy quantum devices, where we try to engineer entangled states within the relevant coherence time 
\cite{knillQuantumComputingVery2005,ciracGoalsOpportunitiesQuantum2012,altmanQuantumSimulatorsArchitectures2021,
flanniganPropagationErrorsQuantitative2022,daleyPracticalQuantumAdvantage2022}.
In systems with local interactions, this spreading is constrained by Lieb-Robinson bounds 
\cite{liebFiniteGroupVelocity1972,bravyi2006liebrobinson,nachtergaele2010lieb,Cheneau2012lightcone,robertsLiebRobinsonBoundButterfly2016}, and the coherence time polynomially restricts the system size over which useful entanglement can be generated. 
However, long-range and non-local interactions can lead to a breakdown of corresponding lightcones for information spreading \cite{hastingsSpectralGapExponential2006,nachtergaele2006liebrobinson,bentsenFastScramblingSparse2019,elseImprovedLiebRobinsonBound2020,tranLiebRobinsonLightCone2021}, up to a conjectured fast scrambling limit \cite{sekino2008fast,lashkari2013towards,hosurChaosQuantumChannels2016,maldacenaBoundChaos2016,
Gharibyan2019,bentsenTreelikeInteractionsFast2019,bentsenFastScramblingSparse2019}. A fast scrambler develops entanglement up to the Page limit \cite{page1993average,foongProofPageConjecture1994} on a timescale that grows only logarithmically, $\sim \log N$ with the system size $N$ \cite{sekino2008fast,maldacenaBoundChaos2016}. The nature of scrambling dynamics on models which interpolate between slow (local) and fast (nonlocal) scrambling regimes are now being explored in both theory 
\cite{kitaevSimpleModelQuantum2015,hosurChaosQuantumChannels2016,swingleMeasuringScramblingQuantum2016,yao2016interferometric,Nahum_2017,nahum2018operator,pappalardiScramblingEntanglementSpreading2018,bentsenTreelikeInteractionsFast2019,marino2019cavity,belyansky2020minimal,xuAccessingScramblingUsing2020,li2020fast,hashizumeTunableGeometriesSparse2022,baoFiniteTimeTeleportation2022} and experiments \cite{Garttner2017measuring,Landsman2019verified,joshiQuantumInformationScrambling2020,Periwal_2021,mi2021information}.

In this work, we identify the onset of scrambling at early times as a dynamical transition that can be characterised via the tripartite mutual information \cite{kitaevTopologicalEntanglementEntropy2006,hosurChaosQuantumChannels2016,iyodaScramblingQuantumInformation2018,gullans2020dynamical,zabalo2020critical,hashizumeTunableGeometriesSparse2022}, which exhibits finite-size scaling collapse around a critical point as a function of a coupling range parameter. We study randomly coupled quantum circuit models with tunable interactions, which can be continuously tuned between the two regimes of different dynamical behaviour. This transition connects to practical applications in noisy devices, especially in identifying regimes where resource states can be generated on timescales that grow only logarithmically with the size of the system, so that the relevant system size can grow exponentially with the coherence time (see, e.g., the generation of entangled resource states in Ref. \cite{baoFiniteTimeTeleportation2022}). Below we analyse this transition for both deterministic and random quantum circuit models, and demonstrate that this transition could be observed with cold atoms in optical tweezer arrays and moving atoms \cite{beugnon2007two-dimensional,hashizumeDeterministicFastScrambling2021,Bluvstein_2022}.

\begin{figure}[htb!]
   \centering
\includegraphics[width=1.0\columnwidth]{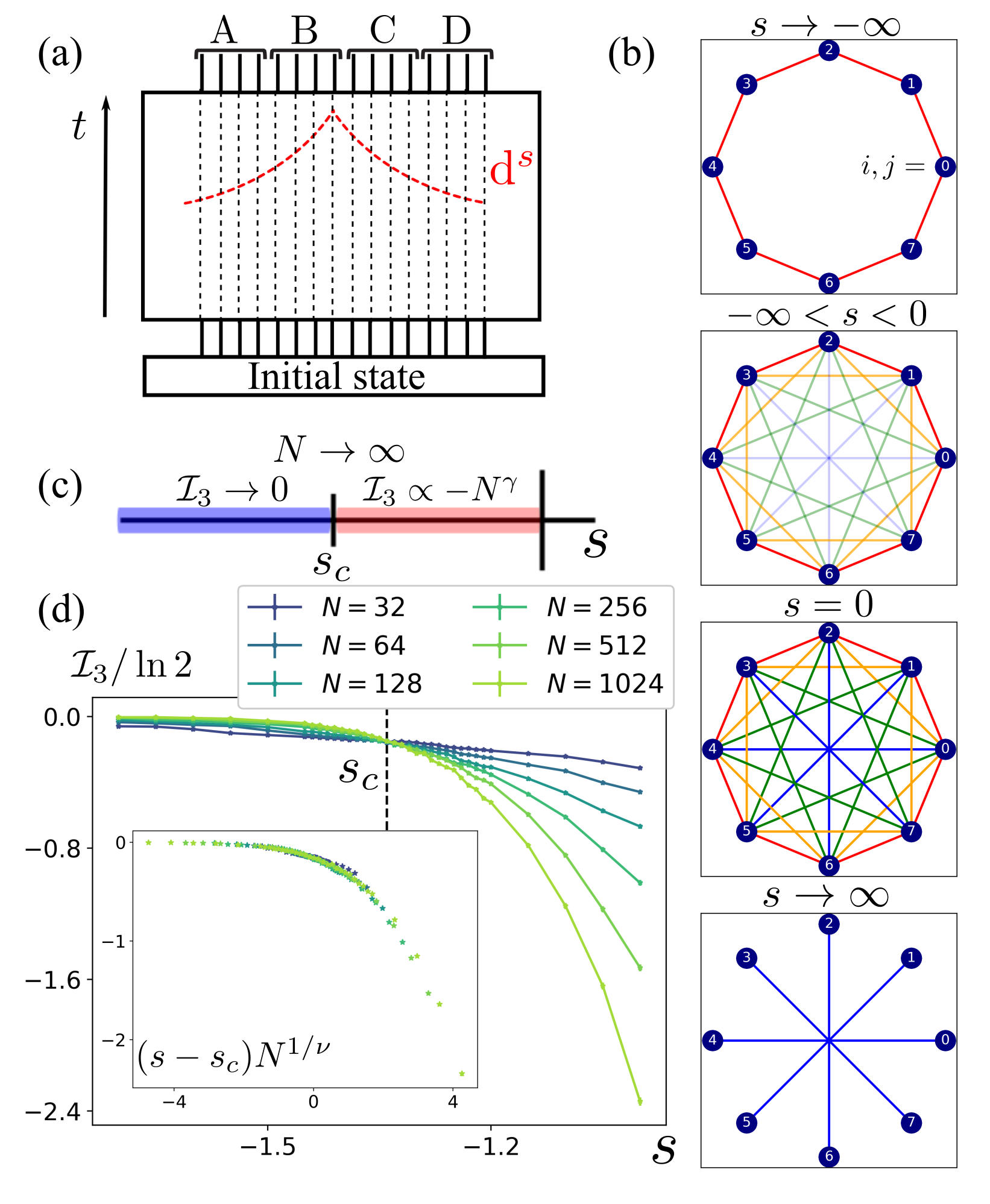}
\caption{
   Onset of scrambling in tunable-range quantum circuits. (a) A linear chain of $N$ qubits $i,j = 0, \ldots, N-1$ (dotted black) evolves in time $t$ under stochastic application 
   of single- and two-qubit Clifford gates $\mathcal{Q}_{ij}$ whose probability distribution $p(d,s) \propto d^s$ (dotted red) depends on distance $d=|i-j|$ with tunable power-law exponent $s < 0$. Output qubits are divided into four contiguous regions $A,B,C,D$ of equal size $N/4$. 
   (b) Tuning the parameter $s$ generates different coupling graphs between qubits. When $s \rightarrow -\infty$ the model is comparable to a nearest-neighbor model, and when $s=0$, all the qubits are coupled randomly with equal probability.      
   (c) Numerical studies of the tripartite mutual information $\mathcal{I}_3$ of three contiguous sub-regions $A,B,C$ as a function of $s$ reveal a crossing at the critical point $s_c$. When $s<s_c$, $\mathcal{I}_3 \rightarrow 0$ (blue) in the thermodynamic limit, indicating an uncorrelated regime. By contrast, when $s>s_c$, the tripartite mutual information is large and negative $\mathcal{I}_3 < 0$ (red), characterizing a scrambled regime. (d) Tripartite mutual information between three output regions $A$, $B$, and $C$ as a function of the tunable parameter $s$ for various systems sizes $N=32,..,1024$ at a fixed time step $t=1$. (Inset) Scaling ansatz for $\mathcal{I}_{3}$ gives $\nu=2.72 \pm 0.52$ and $s_{c}=-1.33$. The numerical results are averaged over 5000 random circuit realizations. 
\label{fig:overview}
}
\end{figure}

 We study this phenomenon in random quantum circuits, which efficiently scramble quantum information and can be treated using both numerical and analytical methods. We first identify the onset of scrambling in numerical simulations of random stochastic Clifford circuits, and then explain our findings using a related Brownian circuit whose scrambling dynamics can be mapped onto a long-range Ising model for a special choice of parameters. Specifically, we consider quantum circuits in 1+1D where qubits $i,j=0,1,..N-1$ are arranged in a linear chain and interact via random two-qubit gates that depend on the linear distance $d=|i-j|$ between them. A tunable power-law exponent $s < 0$ controls the decay of the couplings with distance as shown in \figref{fig:overview}(a) (red-dashed line) \footnote{We use the tunable parameter $s$ instead of the more conventional power-law exponent $\alpha \equiv -s$ to maintain notational consistency with the sparse models studied at the end of this paper.}. 
 
 In these circuits we apply two-qubit gates $Q_{ij}$ between qubits $i$ and $j$ with probability
    \begin{align}
        p(|i-j|,s)=\mathcal{J}|i-j|^{s},
        \label{Eq:Main_Prob_coupling}
    \end{align}
where the normalization factor $\mathcal{J}$ ensures that on average one gate is applied per site during each time step $t=0,1,2,\ldots$. Therefore, regardless of $s$, 
the average number of gates applied per qubit is constant at a given time step $t$. 
For sufficiently short-range interactions $s \rightarrow -\infty$, the qubits mimic a nearest-neighbour (NN) model. The dynamics in this case are governed by the underlying conventional linear light cone constrained by Lieb-Robinson bounds. By contrast, when $s \rightarrow \infty$, terms acting at the longest available distances dominate, as illustrated in \figref{fig:overview}(b). In between these two regimes, when $s=0$, all the qubits are coupled with equal probability. Here it is known that system-wide entanglement builds up on much shorter timescales $t \sim \log N$, logarithmic in system size $N$, allowing for fast scrambling \cite{sekino2008fast,lashkari2013towards,bentsenFastScramblingSparse2019,bentsenTreelikeInteractionsFast2019,li2020fast,belyansky2020minimal,hashizumeDeterministicFastScrambling2021,hashizumeTunableGeometriesSparse2022}.

Our work is concerned with scrambling dynamics in these tunable-range systems at early times, well before the system has achieved volume-law entanglement. To study the growth of entanglement in these models, we measure the tripartite mutual information between three subregions $A, B, C$ of the output qubits:
\begin{align}
    \mathcal{I}_3 \equiv  I(A:B:C)=I(A;B)+I(A;C)-I(A;BC),
    \label{Eq:TMI}
\end{align} 
where $I(A;B)=S^{(2)}_A+S^{(2)}_B-S^{(2)}_{AB}$ is the mutual information between subregions $A$ and $B$. In this paper we will focus primarily on Clifford circuits, where the second Renyi entropy $S^{(2)}_A=-\ln{\mathrm{Tr}[\rho_{A}^{2}]}$ is sufficient to completely characterize the system's entanglement spectrum \cite{renyi1961measures,muller_lennertQuantumRenyiEntropies2013}. 

The tripartite mutual information $\mathcal{I}_3$ vanishes when the regions $ABC$ are uncorrelated, as the amount of information subregions $B$ and $C$ have about $A$ is exactly equal to the information the combined region $BC$ have about $A$. This is because the quantum information is localized. 
However, when quantum correlations have spread across the system, $I(A;BC)>I(A;B)+I(A;C)$ and $\mathcal{I}_3 < 0$. This means that the information contained in the quantum state is delocalized across all three regions $A$, $B$, $C$ and reconstruction of this information requires access to all three regions. Hence, the negativity of the tripartite mutual information serves as a natural measure of many-body entanglement in the system \cite{hosurChaosQuantumChannels2016,gullans2020dynamical,zabalo2020critical}.
Crucially, note that Bell pairs (bipartite entanglement) shared between qubits are not enough to generate negativity in $\mathcal{I}_3$ (\appref{APP:Heuristic_toy}).

Characterizing entanglement generation in the circuit models using $\mathcal{I}_3$ for different values of the tunable parameter $s$ reveals two distinct regimes as shown in \figref{fig:overview}(c). For short-range interactions $s < s_c$ and fixed time step $t$, the tripartite mutual information vanishes $\mathcal{I}_3 \sim 0$ in the thermodynamic limit because there are not enough non-local gates to scramble information throughout the system (blue). On the other hand, for sufficiently long-range interactions $s > s_c$ and fixed time step $t$, the tripartite mutual information becomes large and negative $\mathcal{I}_3 < 0$, indicating that information has been scrambled throughout the entire system (red). At an intermediate value $s = s_c$ between these two regimes we observe signatures of a critical point, which we qualitatively explain in subsequent sections using a pair of toy models. In this sense, our work identifies a dynamical transition in the early-time dynamics of unitary scrambling circuits as a function of the tunable parameter $s$.
\section{Weighted Random all-to-all (W\lowercase{r}AA) model}
Our primary evidence for this transition originates in a family of Clifford circuits 
\cite{Gottesman_1998,Aaronson_2004,selingerGeneratorsRelationsNqubit2015,Nahum_2017,li2021statistical}
that interpolate between the nearest neighbor model and random all-to-all regimes with a tunable parameter $s$. 
Consider a stochastic power-law Clifford circuit acting on $N$ qubits indexed by $i,j=0,1,..N-1$. In each circuit layer, the qubits are randomly paired up, and a random two-site gate from the Clifford group is applied with probability $p(d,s)$ on each qubit pair,
where 
\begin{align}
\label{Eq:Probability_RAA}
   p(d,s)=\normfacWrAA d^s.
\end{align}
Here $d=\min\{N-|i-j|,|i-j|\}$ is the inter-qubit distance with periodic boundary conditions.
Using the definition of the normalization factor discussed after  \eqeqref{Eq:Main_Prob_coupling}, $1/\normfacWrAA=(N/2)^s+2\sum_{d=1}^{N/2-1}d^s$ ensures that each site has only one gate applied to it at one time step $t$. This is achieved after $N-1$ layers.

To characterize the onset of scrambling in this system, we calculate the tripartite mutual information $\mathcal{I}_3$ of three contiguous regions $A$, $B$ and $C$ of size $N/4$ in the output state, across system sizes $N=32, 64, \ldots, 1024$, for different values of the tunable parameter $s$ initialized in the $z$-polarized state. At fixed time step $t=1$, we observe a transition from the slow scrambling regime to the fast scrambling regime at a critical value of $s = s_c$ as shown in \figref{fig:overview}(d). 
The critical point $s_{c}$ is located by looking for a crossing as we vary the system size $64\leq N \leq 1024$. 
Near the transition, we observe that the tripartite mutual information exhibits a scaling collapse as shown in \figref{fig:overview}(d). We find empirically that the data collapse is well-described by the ansatz
\begin{align}
    \mathcal{I}_3=N^{\zeta/\nu}f(|s-s_c|N^{1/\nu}),
\end{align}
where $f$ is a universal scaling function, $\nu$ is the critical exponent of the correlation length $\eta$ and $\zeta=0$
(\appref{APP:FSS}). 
We observe that the data collapses down to universal curves using the resulting estimates of $s_c=-1.33$ and $\nu=2.72 \pm 0.52$ in the inset of \figref{fig:overview}(d). We also check the robustness of this transition for different sub-system sizes in \appref{APP:Robustness_transition}.

\section{Brownian circuit model}
\begin{figure}
    \centering
    \includegraphics[width=0.95\columnwidth]{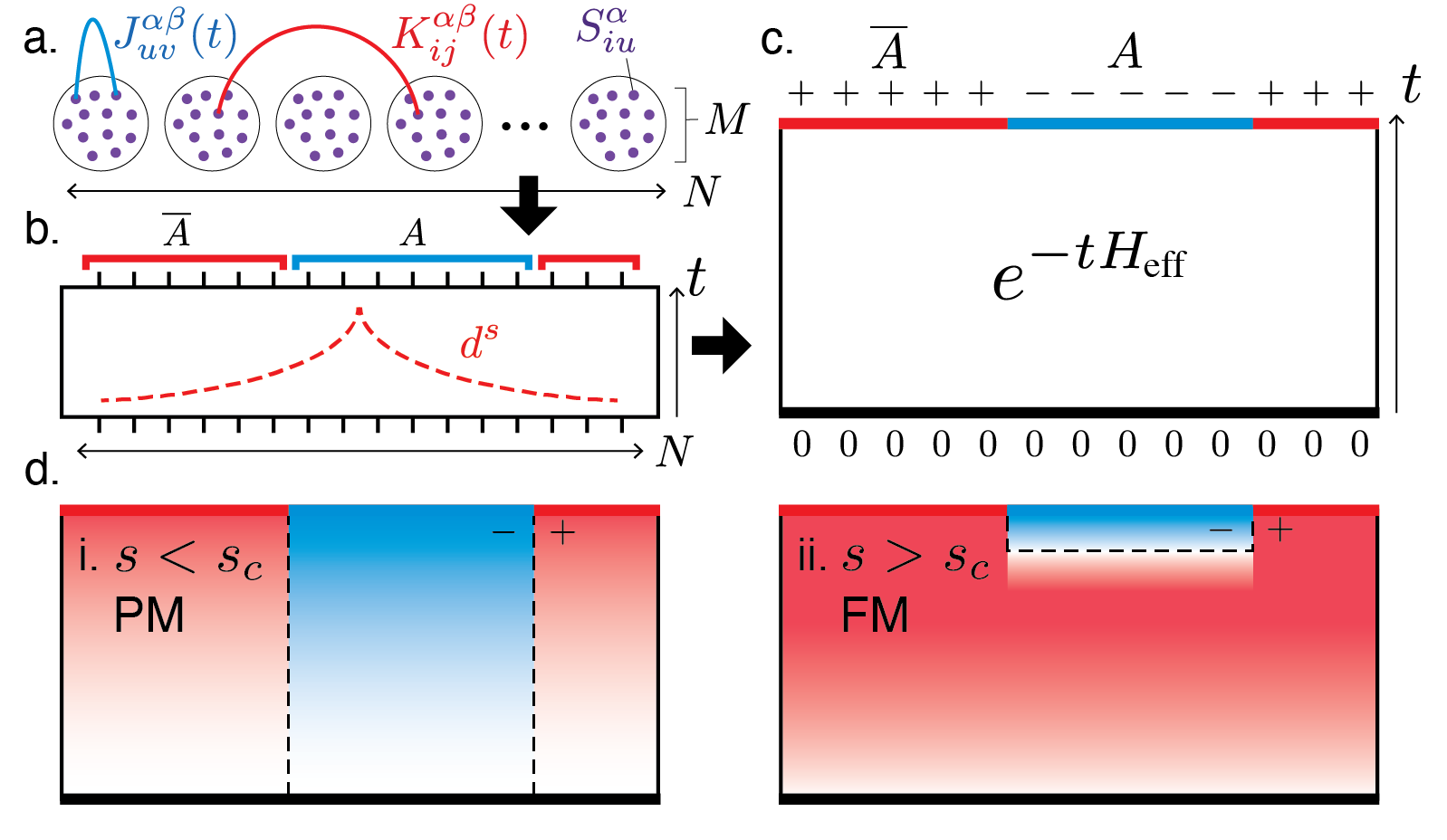}
    \caption{Onset of scrambling as a domain-wall depinning transition in a Brownian quantum circuit (a). Averaging over the Brownian couplings $J_{uv}^{\alpha \beta}(t),K_{iu,jv}^{\alpha \beta}(t)$ and taking the large cluster size limit $M \rightarrow \infty$ maps the calculation of the Renyi entropy $S_A^{(2)}$ (b) of a subregion $A$ onto a quantum statistical mechanics problem (c) with effective Hamiltonian $H_{\mathrm{eff}}$ and exactly two saddle point solutions $\ket{\psi_{\pm}}$ (red, blue). For short-range interactions $s < s_c$ (d.i), the system is in a high-temperature paramagnetic (PM) phase where it is energetically favorable to form vertical domain walls (d.i., dotted black) that are pinned to the bottom of the circuit. For $s > s_c$ (d.ii.), the system is in a low-temperature ferromagnetic (FM) phase where vertical domain walls are energetically expensive and it is therefore favorable to form a horizontal domain wall (d.ii., dotted black) that is `depinned' from the bottom boundary.}
    \label{fig:domainwalls}
\end{figure}

We can understand these results using a closely related Brownian circuit model \cite{lashkari2013towards,shenkerStringyEffectsScrambling2015,Chen2019,sunderhaufQuantumChaosBrownian2019,xuLocalityQuantumFluctuations2019,zhouOperatorDynamicsBrownian2019,zhou2020operator,bentsen2021measurement,sahu2022entanglement}. The utility of this model is that the growth of entanglement can be mapped onto a statistical mechanics problem. Further, for a special choice of parameters the dynamics can be mapped onto a long-range Ising model whose thermodynamics is well-understood \cite{dyson1969existence,fisher1972critical,cannas1995onedimensional}. For any given WrAA model, we construct a related Brownian circuit that acts on $N$ clusters of $M$ spins arranged in a line as shown in \figref{fig:domainwalls}a, where $i,j = 0,\ldots,N-1$ label the clusters and $u,v = 0,\ldots,M-1$ label the spins $\vec{S}_{iu} = S_{iu}^{\alpha}$ within each cluster. During each infinitesimal timestep $dt$, the system evolves under the unitary operator
\begin{align}
    U_t &= e^{-i H(t) dt} \\
    &= \exp \left[- i J_{uv}^{\alpha \beta}(t) S^{\alpha}_{iu} S^{\beta}_{iv} dt - i K_{iu,jv}^{\alpha \beta}(t) S^{\alpha}_{iu} S^{\beta}_{jv} dt \right] \nonumber
\end{align}
where repeated indices are implicitly summed over. Here the Brownian couplings $J_{uv}^{\alpha \beta}(t)$ generate intra-cluster interactions while the couplings $K_{iu,jv}^{\alpha \beta}(t)$ generate inter-cluster interactions. These couplings are white-noise variables with zero mean and variance
\begin{align}
    \mathbb{E} \left[ J_{uv}^{\alpha \beta}(t) J_{u'v'}^{\alpha' \beta'}(t') \right] &= \frac{\mathcal{J}}{M dt} \left(1 - b \right) \delta^{\alpha \alpha'} \delta^{\beta \beta'} \delta_{uu'} \delta_{vv'} \delta_{t t'} \\ \nonumber
    \mathbb{E} \left[ K_{iu,jv}^{\alpha \beta}(t) K_{i'u',j'v'}^{\alpha' \beta'}(t') \right] &= \frac{\mathcal{J}}{M dt} b A_{ij} \ \delta^{\alpha \alpha'} \delta^{\beta \beta'} \delta_{ii'} \cdots \delta_{t t'}, \nonumber
\end{align}
where $\mathcal{J}$ controls the overall coupling strength, $b$ controls the relative strength of the intra- and inter-cluster couplings, and $A_{ij} = A \magn{i-j}^s$ is the normalized inter-cluster coupling matrix falling off as a power law with exponent $s$, with normalization factor $1/A = \sum_i \magn{i}^s$.

We are interested in the tripartite mutual information $\mathcal{I}_3$ of this system after a fixed evolution time $t$, which requires us to compute entanglement entropies $S_A,S_B,S_{AB},\ldots$ of various subregions of the system at the final time $t$ as shown in \figref{fig:domainwalls}b. For simplicity, we focus here on computing the second Renyi entropy $S_A^{(2)}$, which is equivalent to introducing two copies of the system and measuring the expectation value of the SWAP operator (`doing a SWAP test’) on the region $A$. Performing the disorder average over the random Brownian couplings $J_{uv}^{\alpha \beta}(t),K_{iu,jv}^{\alpha \beta}(t)$, and taking the limit $M \rightarrow \infty$ 
(\appref{app:Brownian}), one can show that the Renyi entropy reduces to a statistical mechanics problem
\begin{equation}
    e^{-S_A^{(2)}} \propto \bbra{\Psi_T} e^{-t H_{\mathrm{eff}}} \kket{\Psi_0}
\end{equation}
where $H_{\mathrm{eff}}$ is an effective ferromagnetic Hamiltonian with exactly two ground states $\ket{\psi_{\pm}}$ (red and blue regions in \figref{fig:domainwalls}). Here, the total time acts as an inverse temperature $t = \beta$ (note that $t$ is not the temperature.) Thus the entropy $S_A^{(2)}$ is governed by the Euclidean (thermal) propagator between the initial state $\ket{\Psi_0}$ and the final state $\ket{\Psi_T} = \ket{\psi_-}_A \ket{\psi_+}_{\overline{A}}$ as illustrated in \figref{fig:domainwalls}c. This mapping to an effective statistical mechanics model is similar to the mappings one finds in random quantum circuit models \cite{nahum2018operator,vonKeyserlingk2018operator,jian2020measurement,bao2020theory,nahum2021measurement,li2021statistical,baoFiniteTimeTeleportation2022,potterEntanglementDynamicsHybrid2022,fisher2023random} and random tensor network models \cite{li2021statistical,haydenHolographicDualityRandom2016}; these mappings are fundamentally related to Schur-Weyl duality \cite{li2021statistical}.


In this language, the transition in $\mathcal{I}_3 $ can be understood as an ordering transition in the statistical mechanics model $H_{\mathrm{eff}}$ as illustrated in \figref{fig:domainwalls}d. For sufficiently short-range interactions $s < s_c$ (at fixed total time $t$), the system is in its disordered high-temperature phase. In this regime, the short-range interactions are too weak to stabilize a ferromagnetic phase at temperature $1/t$. In other words, domain walls are cheap, and so the energy gap $\Delta$ of $H_{\mathrm{eff}}$ is smaller than the temperature $1/t$. Therefore the bulk is in a high-temperature paramagnetic (PM) phase where the system has not yet reached thermal equilibrium.
In this case, the free energy is dominated by `vertical' domain walls that extend straight down from the region $A$ and are `pinned' to the bottom edge of the circuit, shown in dotted black in \figref{fig:domainwalls}d.i. As a result, we find $S_A^{(2)} = 2g(t, s)$ for some function depending only on time $t$ and exponent $s$, and not on the
subregion size $\magn{A}$, such that $\mathcal{I}_3 = 0$.

By contrast, for sufficiently large $s > s_c$, the long-range couplings make domain walls more costly and the energy gap $\Delta$ correspondingly increases. In this regime, the long-range interactions are strong enough to stabilize a low-temperature ordered ferromagnetic (FM) phase as shown in \figref{fig:domainwalls}d.ii. Whereas the bulk quickly finds the ground state $\ket{\psi_+}$ favored by the majority boundary region $\overline{A}$, the boundary condition $\ket{\psi_-}$ on region $A$ induces a horizontal domain wall (dotted black) that is `depinned' from
the bottom edge. In this case we find volume-law entanglement entropy $S_A^{(2)} = c \magn{A}$ for some fixed constant $c$, and correspondingly negative tripartite mutual information $\mathcal{I}_3  < 0$. The transition between these two regimes occurs when the bulk `freezes' into its ordered low-temperature phase as a result of increasingly strong nonlocal interactions.

In this sense, the tripartite mutual information $\mathcal{I}_3$ probes a domain-wall depinning transition in the Brownian circuit that is driven by an ordering transition in the effective Hamiltonian $H_{\mathrm{eff}}$. The reason $\mathcal{I}_3$ diagnoses this transition is precisely because the contributions from vertical domain walls separating neighboring regions cancel (\figref{fig:domainwalls}d.i., dotted black) while contributions from horizontal domain walls do not cancel \cite{zabalo2020critical,Li2023}. Thus, $\mathcal{I}_3 = 0$ unless the horizontal domain wall (\figref{fig:domainwalls}d.ii., dotted black) has `depinned' from the bottom edge. For sufficiently long-range interactions $s > s_c$ the tripartite mutual information is negative precisely because the input and output of the circuit are separated by this depinned domain wall.

While the preceding discussion has been largely heuristic, we can conclusively demonstrate this ordering transition for a special choice of parameters by mapping the dynamics of $H_{\mathrm{eff}}$ onto a long-range Ising model. In particular, consider the above Brownian cluster model in a limit with weak inter-cluster couplings $b \ll 1$ and strong intra-cluster couplings $\mathcal{J} t \gg 1$ such that $b \mathcal{J} t$ is a fixed number of order one. This is equivalent to strong scrambling within each cluster and weak scrambling between the clusters, which is the regime our Clifford circuits probe. In this limit, we can show (Appendix \ref{app:Brownian}) that each cluster $i$ collectively behaves as an Ising spin $\sigma_i = \pm 1$, where the spin labels the two possible saddle-point solutions $\ket{\psi_{\pm}}$. In this region of parameter space, the boundary conditions $\lvert \psi_{0,t} \rrangle$ on each cluster $i$ act as effective magnetic fields $h_i$, where $h_i = \pm1$ for $i \in \overline{A},A$. The Brownian circuit in this limit is therefore equivalent to a long-range Ising model, whose ordering transition is well understood \cite{dyson1969existence,fisher1972critical,cannas1995onedimensional}. In fact, the critical exponents of the long-range Ising model are known to be mean-field and numerically agree with the critical exponents we find in our Clifford simulations, suggesting that the transition in question is itself mean field. Specifically, the mean-field critical exponent for the long-range Ising model predicted in \cite{fisher1972critical} is, in terms of our parameters, $\nu = -1/(1+s_c) \approx 3$, which agrees with our Clifford results to within a single standard deviation. It would be interesting to put this conjecture on a stronger footing in future studies.


One other heuristic toy model that we devised to understand the approach to this transition is to calculate the probability of application of two qubit gates to a single qubit that crosses both the boundaries of a given subregion. These gates are responsible for the proliferation of entanglement entropy in the given quantum system 
(\appref{APP:Heuristic_toy}). Such a simple model, however, is not sufficient to account for the negativity of tripartite mutual information, as a collection of EPR pairs always has $\mathcal{I}_3 = 0$. In order to have negativity $\mathcal{I}_3 < 0$, the system must have multi-party entanglement, which is present in the Clifford circuit and Brownian model, but not in this simple heuristic model.

\section{Towards experimental observation}
Having explored the onset of scrambling numerically in the WrAA Clifford circuit and analytically in a related Brownian circuit model, it is interesting to ask whether this transition can be observed in experiments. Tunable power-law interactions of the type studied above are naturally accessible in trapped ion experiments, and the tripartite mutual information can be measured in principle by interfering many-body twins \cite{daley2012measuring,islam2015measuring} or by performing randomized measurements \cite{brydges2019probing, elben2022randomized}. Here we pursue a slightly different angle and ask whether the same phenomenon appears in systems with sparse interactions that can be engineered in ensembles of Rydberg atoms with optical tweezers. 
Building on ideas proposed in \cite{hashizumeDeterministicFastScrambling2021} we demonstrate that the transition studied above can be observed in near-term experiments using optically trapped neutral Rydberg atoms  \cite{Saffman_2010_Quantum_info_ryd, Adams_2020, Henriet_2020, Morgado_2021}. Nonlocal couplings in the system are generated by a quasi-1D shuffling process employing optical tweezers that rearrange atomic positions \cite{beugnon2007two-dimensional,barredo2016atom-by-atom, endres2016atom-by-atom,barredo2018synthetic,Bluvstein_2022}. Each rearrangement executes a ``Faro Shuffle'' which moves the atom originally located at lattice site $i$ to lattice site $i'$ according to the map \cite{aldous1986shuffling,diaconis1983mathematics}  
\begin{align}
    i'=\mathcal{R}(i=b_{m}...b_{2}b_{1})=b_{1}b_{m}...b_{2}.
    \label{Eq:Faro_Shuffle}
\end{align}
This nonlocal mapping permutes the bit order of the atomic index $i=b_{m}...b_{2}b_{1}$, written in binary such that the least significant bit $b_{1}$ of $i$ becomes the most significant bit of $i'=\mathcal{R}(i)$. 

\begin{figure}[htbp]
\includegraphics[width=1\columnwidth]{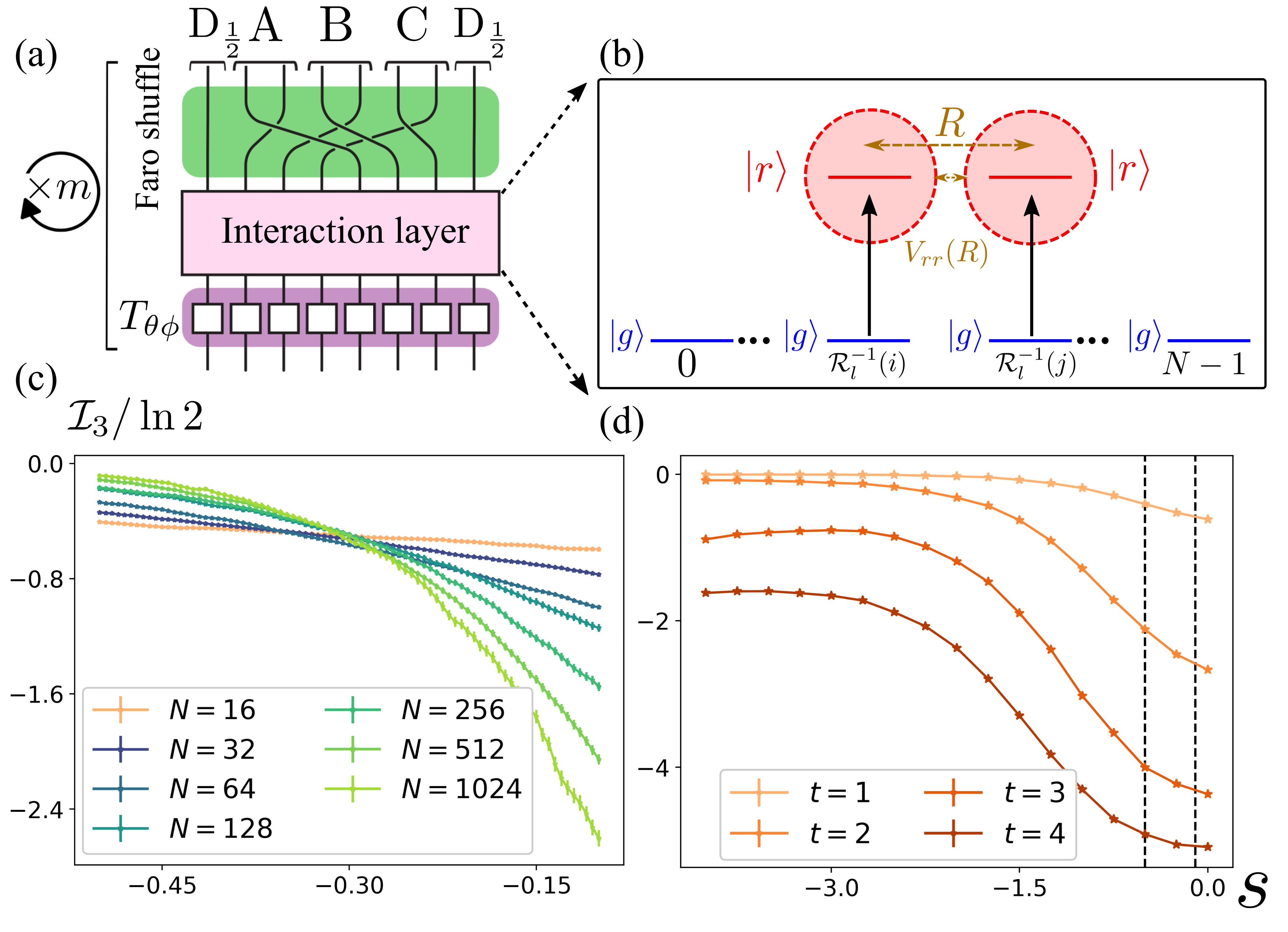}
\caption{Onset of scrambling in a deterministic Riffle circuit with random initial state (a) Our circuit implementation, is composed of inverse Faro-shuffles (green), global rotations (purple), and an interaction layer (pink) composed of weighted $CZ_{even}^{w}$ and $CZ_{odd}^{w}$ gates. The Controlled-Z ($CZ$) gates between neighboring atoms separated by an inter-atomic distance of $R$ can be realized using strong van der Waals interactions $V_{rr}(R)$ between Rydberg states (b). The application of random two-qubit gates is achieved using classical random sampling to select qubits $i,j$ according to the probability distribution \eqeqref{eq:probcoupling_det}. The chosen qubits have a corresponding atomic positions given by $\mathcal{R}^{-1}_{l}(i)$ and $\mathcal{R}^{-1}_{l}(j)$, where $l$ is the number of inverse shuffle operations. They are then excited from the ground state $\ket{g}$ to the Rydberg state $\ket{r}$ where they interact via strong Rydberg Rydberg interactions (RRI).  
(c). Tripartite mutual information ($\mathcal{I}_3$) between three regions $A$, $B$, and $C$ in the bulk as a function of the tunable parameter $s$ for various systems sizes $N=32,64,..1024$. This transition is observed for system sizes as small as $N=16$ (orange) at $t \sim O(1)$. (d) $\mathcal{I}_3$ between three regions $A$, $B$, and $C$ as a function of the tunable parameter $s$ for $N=16$ for different values of time steps $t$ (light to dark orange). Black dotted vertical lines highlight the window over which $\mathcal{I}_3$ is plotted as a function of $s$ in (c).  The numerical results are averaged over 3000 random circuit realizations.
\label{Fig:Det_Circuits}
}
\end{figure}

Repeated shuffling operations lead to a dramatic rearrangement of the atomic positions and the resulting nonlocal couplings can rapidly generate many-body entanglement. The coupling operations in the circuit occur in the interaction layer as shown in \figref{Fig:Det_Circuits}(a) and are achieved using stochastically applied alternating even and odd layer of Controlled-Z (CZ) gates. Combining this, along with global rotations, we implement a strong scrambling circuit ($\mathrm{sc}$)
\begin{align}\label{Eq:Det_Circuit} 
    \mathcal{D}_{\mathrm{sc}} &= \mathcal{E}_{\mathrm{sc}}^{m}\mathcal{O}_{\mathrm{sc}}^{m}, 
\end{align}
where 
\begin{align}\label{Eq:Even_Circuit}
    \mathcal{E}_{\mathrm{sc}}&=[\mathcal{R}^{-1} \mathrm{CZ}_{even}^{w}\mathrm{T}_{\theta \phi}]
\end{align}
is the even circuit iteration, and
\begin{align}
    \mathcal{O}_{\mathrm{sc}}&=[\mathcal{R}^{-1} \mathrm{CZ}_{odd}^{w}\mathrm{T}_{\theta \phi}]
\end{align}
the odd circuit iteration, with $\mathrm{T}_{\theta \phi}=\mathrm{H}\mathrm{P}$, a combination of global Hadamard $\mathrm{H}$ and global phase gates $\mathrm{P}$.
The weighted $CZ_{even}^{w}$ and $CZ_{odd}^{w}$ gates couple qubits $i < j$ with probability $p(d,s)$ given by
\begin{equation}
   p(d,s) =
     J_{\mathcal{D}_s} d^s.
  \label{eq:probcoupling_det}
  \end{equation}
The normalization factor $J_{\mathcal{D}_s}$ ensures that one gate, on average, is applied per site during each timestep $t$. In our sparse coupling circuit, this is achieved after $m=\mathrm{log}_{2}N$ even $\mathcal{E}_{s}$ and $m=\mathrm{log}_{2}N$ odd $\mathcal{O}_{s}$ circuit iteration. One even circuit iteration $\mathcal{E}_{\mathrm{sc}}$ is composed of globally applied $\mathrm{T}_{\theta \phi}$, followed by a stochastic random application of $CZ_{even}^{w}$ gates and a ``Faro Shuffle'' as illustrated in \figref{Fig:Det_Circuits}(a). The same procedure applies to one odd circuit iteration $\mathcal{O}_{\mathrm{sc}}$. 

Similar to \figref{fig:overview}(d), we observe the dynamical transition in \figref{Fig:Det_Circuits}(c), characterized by the negativity of the tripartite mutual information of three contiguous regions $A$, $B$, and $C$ across system sizes $N= 16,32, 64, ..1024$, at a fixed time step $t=1$ initialized in a completely random state. The three contiguous regions $A$, $B$, and $C$ of the output set of qubits are chosen in the bulk as shown in \figref{Fig:Det_Circuits}(a) to avoid any boundary effects on the calculations. A random initial state is characterized by an arbitrary polarization ($x$, $y$ or $z$) qubit state at each site. We also perform finite-size scaling analysis for this model in \appref{APP:Scaling_collapse_Det_model}.
From the perspective of near-term experimental realization, this transition can be observed in system sizes as small as $N=16$ and $N=32$, for a deterministic initial state, avoiding individual addressing of the qubit states. Furthermore, with increasing time steps, the many-body entanglement generated in the system increases and is characterized by an appreciable negative value of $\mathcal{I}_3$ 
(\appref{APP:Transition_depths}). We have demonstrated in \figref{Fig:Det_Circuits}(d) the change in the value of $\mathcal{I}_3$ as a function of the tunable parameter $s$ for different values of time steps $t$ for $N=16$.   

To explore this dynamical transition in experiments, we propose to use a long-lived ground state $\ket{g}$ and a short-lived excited Rydberg state $\ket{r}$ as our qubit states. A random initial state can be prepared using classical random sampling and rotating individual spins constituting the qubit state.
In each of the constituting even $\mathcal{E}_{\mathrm{sc}}$ and odd $\mathcal{O}_{\mathrm{sc}}$ circuit iteration, the implementation of Hadamard and Phase gates can be achieved by single-qubit rotations. For the entangling operations, we randomly sample pairs of qubits $i, j$ according to the probability distribution \eqeqref{eq:probcoupling_det}.  These chosen qubits with the corresponding atomic indices $i'=\mathcal{R}^{-1}_{l}(i)$ and $j'=\mathcal{R}^{-1}_{l}(j)$, with $l$ the number of inverse shuffle operations, are then excited to the Rydberg state $\ket{r_{i'}}$ and $\ket{r_{j'}}$ respectively from the $\ket{g}$ state, and the CZ gates between these qubits can be realized using strong Rydberg-Rydberg interactions (RRI) $V_{rr}$ as illustrated in \figref{Fig:Det_Circuits}(b) \cite{jaksch2000fast,lukin2001dipole,heidemann2007evidence,muller2014implementation,theis2016high,bernien2017probing,isenhower2010demonstration,madjarov2020high}. An alternative way to implement the weighted application of gates in the interaction layer would be to modulate different pulses according to the probability given in \eqeqref{eq:probcoupling_det} in different stages of the inverse shuffling operations. Additionally, we also observe these transitions in other sparse models including the powers-of-two circuit model 
\cite{gubserContinuumLimitsSparse2018,bentsenTreelikeInteractionsFast2019,Periwal_2021,hashizumeTunableGeometriesSparse2022}
(\appref{APP:PWR2_circuit}).

Measuring the tripartite mutual information involves measuring entanglement entropies of different subregions. Clifford circuits have a flat entanglement spectrum, so measuring the second Renyi entropy is equivalent to measuring the von Neumann entropy. In practice the second-order Renyi entropy can be measured in the cold atom setup by quantum interference of many-body twins \cite{daley2012measuring,islam2015measuring} or by performing randomized measurements \cite{brydges2019probing,elben2022randomized}. For $N=16$, we would require to measure the entanglement entropy of a maximum of 8 qubits. For randomized measurements, it is known that the number of measurements required for estimating the second-order Renyi entropy $S_A^{(2)}$ scales exponentially with the size of the subregion $A$ \cite{brydges2019probing,elben2022randomized}. Hence, by preparing a single copy of the qubit state of interest at each time step, the estimation of entanglement entropy can be done using average $10^3$ to $10^4$ measurements. These measurements tend to be quite challenging in practice, however, and it is important to think carefully about how a given protocol would perform given the realities of dissipation and repetition rates in specific platforms.
\section{Summary and Outlook}
In this work we studied a dynamical transition between slow and fast scrambling in quantum circuit models. In circuits where short-range couplings dominate ($s < s_c$), scrambling is constrained by local Lieb-Robinson bounds, leading to slow scrambling behavior. Beyond a critical interaction exponent $s > s_c$, these bounds break down such that information can be scrambled on a much faster timescale $t \sim \log N$. In both cases, we diagnose the presence of scrambling using the negativity of the tripartite information $\mathcal{I}_3 < 0$. We also show that the dynamics of a related long-range Brownian model can be mapped to the long-range Ising model in a particular parameter regime. In particular, we also estimate the mean-field critical exponent and find it to agree with our Clifford simulation results to within a single standard deviation thus supporting the presence of a critical point and hence a phase transition. 

One major outlook would be to probe the transition by teleporting information via the Hayden-Preskill-Yoshida-Yao protocol \cite{hayden2007black,yoshida2017efficient, Yoshida2019Disentangling,Bao_noisy_Hawking_2021, hashizumeDeterministicFastScrambling2021}. For weak scrambling $s < s_c$ we expect teleportation to fail except at timescales comparable to the system size, while for strong scrambling $s > s_c$ we expect it to succeed on timescales $\propto \log{N}$. We leave the characterization of this phenomenon to future work. Further, our studies here focused only on the second Renyi entropy. This is sufficient for our current purposes since Clifford circuits have a flat entanglement spectrum, but for more general circuits one would need to also consider higher-order Renyi entropies. These questions could be addressed using more sophisticated Brownian circuit models that compute higher Renyi entropies. In a similar vein, the phase diagram appears very similar for both the sparse and dense coupling schemes. We likewise observe similar features regardless of whether the model is a deterministic or random scrambler. It would be interesting to understand whether these models actually belong to the same universality class. Data for this manuscript is available at \cite{data_open_access}.

\section{Acknowledgements}
We thank Jon Pritchard for stimulating and helpful discussions. Work at the University of Strathclyde was supported by the EPSRC (Grant No. EP/T005386/1), M Squared Lasers Ltd, the EPSRC Programme Grant DesOEQ (EP/P009565/1), the EPSRC Quantum Technologies Hub for Quantum Computing and simulation (EP/T001062/1), the European Union’s Horizon 2020 research and innovation program under grant agreement No. 817482 PASQuanS, and AFOSR grant number FA9550-18-1-0064. Results were obtained using the ARCHIE-WeSt High Performance Computer (www.archie-west.ac.uk) based at the University of Strathclyde. G.B. is supported by the DOE GeoFlow program (DE-SC0019380).



\bibliography{References.bib}

\newpage 
\appendix

\onecolumngrid

\section{Finite size scaling}
\label{APP:FSS}
For performing the finite size scaling analysis, we use 5000 circuit realizations to locate the critical point $s_{c}$ through the crossing for larger system sizes $64\leq N \leq 1024$.
Using this estimated value of $s_c$, we collapse data according to the scaling ansatz  
    \begin{align}
    \mathcal{I}_3=N^{\zeta/\nu}f(|s-s_c|N^{1/\nu})
    \label{Eq:Supp_Scaling}
    \end{align}
where $f$ is a universal scaling function and $\nu$ and $\zeta$ are the critical exponents of the correlation length $\eta$. The data collapse problem is framed as an optimization problem and the critical exponents including its standard errors are fitted by Nelder–Mead algorithm. These routines are carried out by using the scientific Python package pyfssa \cite{andreas_sorge_2015_35293,autoscale_2009}

\section{Powers-of-two sparse circuit}
\label{APP:PWR2_circuit}
\begin{figure}[htbp]
\includegraphics[width=1.0\columnwidth]{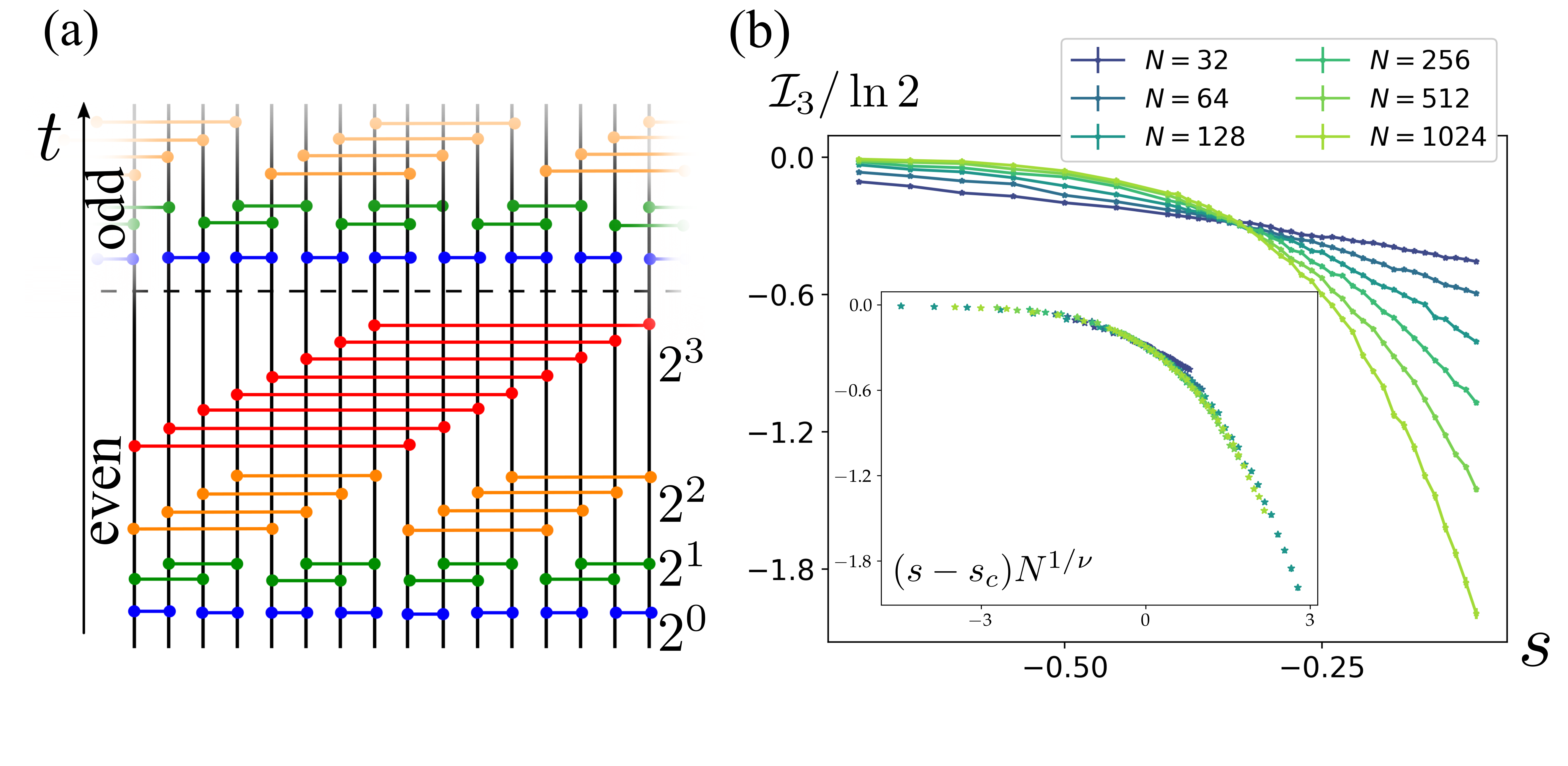}
\caption{\textbf{Dynamical transition in the PWR2 circuit} (a) Qubits arranged linearly in a 1D chain are coupled only if they are separated by a distance equal to an integer power of 2: $\magn{i-j} = 2^{m-1}$ for $m = 1,\ldots,\log_2 N$. The two-qubit Clifford gates are applied stochastically according to the probability distribution Eq. \eqref{eq:probcoupling} in a bricklayer pattern with alternative even and odd interaction layers. (b) Tripartite mutual information $\mathcal{I}_3$ between three regions $A$, $B$, and $C$ as a function of the tunable parameter $s$ for various systems sizes $N=32,..,1024$. (Inset) Scaling collapse of $\mathcal{I}_3$ gives $\nu=2.78 \pm 0.514$ and $s_{c}=-0.33$. The numerical results are averaged over 5000 random circuit realizations.
\label{Fig:PWR2_circuit}
}
\end{figure}
In addition to the Random All-to-All model, and the neutral atoms-based sparse model, we also investigate the powers-of-two (PWR2) circuit model with sparse nonlocal couplings, where random two-qubit gates are applied on sites $i,j=0,\ldots N-1$.
In this circuit, two qubits are coupled if and only if the qubits are separated by an integer powers of $2$: $|i-j|=2^{m-1}$ for $m=1,\ldots,\log_2 N$ as illustrated in \figref{Fig:PWR2_circuit}(a).
The random two-qubit gates $Q_{ij}$ are drawn from the two-qubit Clifford group, 
and they are applied between two qubits in a brickwork pattern. 
The brickwork pattern of the circuit is constructed by distributing the gates with the same distance into layers, with alternating even and odd blocks.
Within each block, the layers corresponding to gate of distance $2^{0}$ to $2^{\log_2 N -2}$ are stacked sequentially, 
where in the even blocks, the gate between qubit $i$ and $j=i+2^{m}\mod N$ is applied when $\lfloor i/2^m \rfloor,2=0$ and $=1$ for the odd blocks.
The layers corresponding to gates of length $2^{\log_2 N-1}$ are placed after one iteration of even and odd block.
In each layer, the gates are applied with the probability $p(|i-j|,s)$, which is dictated by the distance $|i-j|$ and exponent $s$: 
\begin{align}
   p(|i-j|,s) =
   \begin{cases}
      J_{\mathrm{PWR2},s}|i-j|^s &|i-j| =2^{m-1} \\
      0 &$Otherwise$
   \end{cases}
   \label{eq:probcoupling}
\end{align}
Here the normalization factor $1/J_{\mathrm{PWR2},s}=(N/2)^{s}+2\sum_{m=1}^{\log_2 N-1} 2^{(m-1)s} $ is chosen such that in the limit of random unitary circuit, 
the dynamics recover that of an anisotropic XY model on a sparsely coupled graph with Kac normalization  \cite{bentsenTreelikeInteractionsFast2019,baoFiniteTimeTeleportation2022}.
In this formulation, after one iteration of even and odd blocks, on average one gate is applied per qubit. 
Therefore, we define one increment in time step ($t=0$ to $t=1$), as one complete iteration of even and odd block, such that regardless of $s$, 
the average number of gates applied per qubit is constant at a given time step $t$. We observe the dynamical transition between slow scrambling and fast scrambling regime by plotting tripartite mutual information $\mathcal{I}_3$ as a function of the tunable parameter $s$ for various system sizes $N=32,64,..,1024$ as shown in \figref{Fig:PWR2_circuit}(b). We also estimated the critical $s_c$ and the corresponding critical exponent $\nu$ by using the finite size scaling analysis mentioned in \appref{APP:FSS}. The collapsed data estimated using finite size scaling is shown in the inset of \figref{Fig:PWR2_circuit}(b). 
\section{Robustness of the transition to different sub-system sizes}
\label{APP:Robustness_transition}
\begin{figure}
\includegraphics[width=1.0\columnwidth]{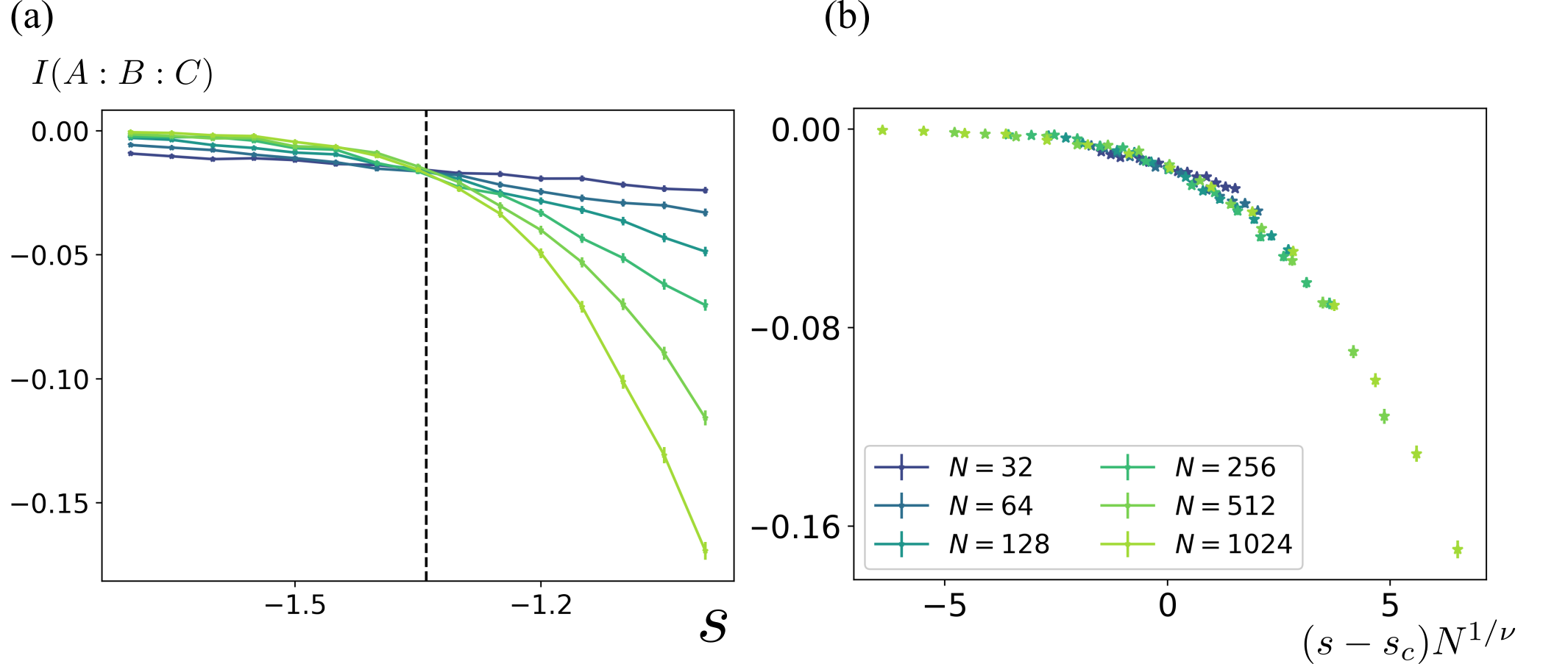}
\caption{ (a) Tripartite mutual information between three output regions $A$, $B$, and $C$ of size $N/8$ as a function of the tunable parameter $s$ for various systems sizes $N=32,..,1024$ for weighted random all to all model at a fixed time step $t=1$. (b). Scaling ansatz for $\mathcal{I}_{3}$ gives $\nu=2.23 \pm 0.67$ and $s_{c}=-1.33$. The numerical results are averaged over 12000 random circuit realizations. 
\label{Fig:Robustness}
}
\end{figure}
We also observe this dynamical transition when the output set of qubits is divided into subregions of size $N/8$. The tripartite mutual information $\mathcal{I}_{3}$ of three contiguous subregions of size $N/8$ in the output state is plotted as a function of the tunable parameter $s$ for the weighted random all to all model across system sizes $N=32, 64, .., 1024$ as shown in \figref{Fig:Robustness}(a). We observe a transition from the slow scrambling to the fast scrambling regime at a fixed time step $t=1$ at a critical value $s_c=-1.33$ same as the analysis done in the main text. This shows that a change in boundary conditions does not affect the bulk physics and we observe the transition at the same point. We also observe scaling collapse as shown in \figref{Fig:Robustness}(b) with critical exponent $\nu=2.23 \pm 0.67$. 
\section{Scaling collapse analysis for the neutral atoms-based sparse model}
\label{APP:Scaling_collapse_Det_model}
\begin{figure}
\includegraphics[width=0.5\columnwidth]{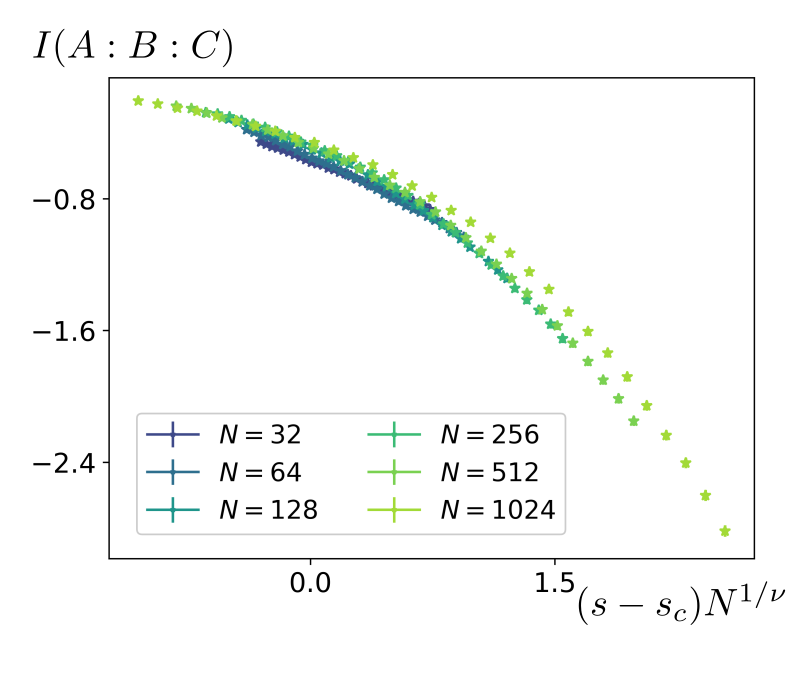}
\caption{Scaling analysis for the neutral atoms-based sparse model with critical exponent $\nu=2.79 \pm 0.39 $ and $s_c=-0.29$. The numerical results are averaged over 5000 random circuit realizations.
\label{Fig:Scaling_collapse_deterministic}
}
\end{figure}

The dynamical transition observed in the neutral atoms-based sparse model allows us to create entangled states with fewer gates. The scaling ansatz for $\mathcal{I}_{3}$ gives $\nu=2.79 \pm 0.39 $ and $s_c=-0.29$ as shown in \figref{Fig:Scaling_collapse_deterministic}. 
\section{Fast scrambling dynamical transition at different time steps}
\label{APP:Transition_depths}
\begin{figure}[htbp]
\centering
\includegraphics[width=1.0\columnwidth]{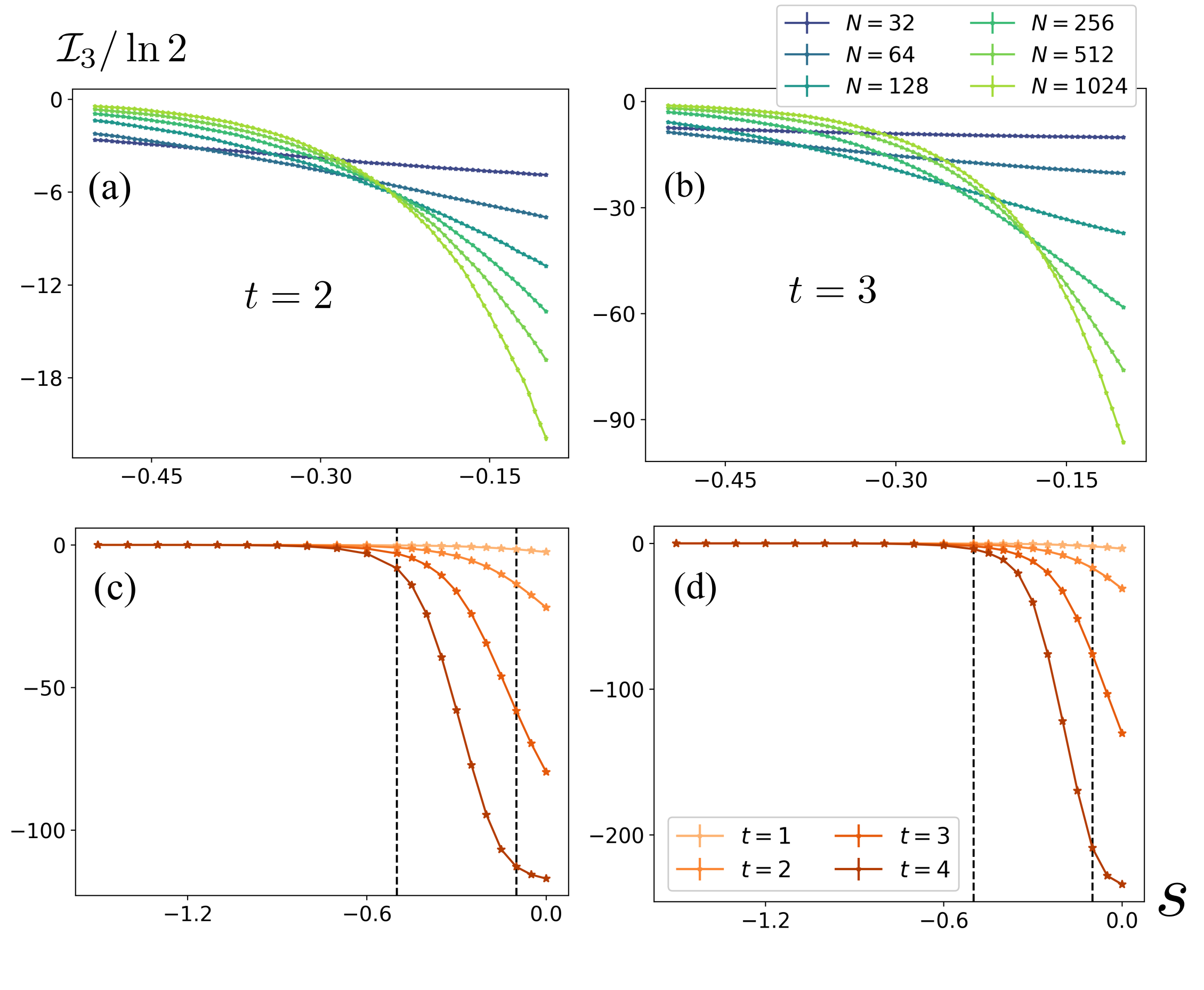}
\caption{\textbf{Fast scrambling dynamical transition at different time steps for the neutral atoms-based model} The Tripartite mutual information $\mathcal{I}_3$ between three regions $A$, $B$, and $C$ of the output set of qubits as a function of the tunable parameter $s$ for various system sizes $N=32,..512$ for (a) $t=2$ and (b) $t=3$. $\mathcal{I}_3$ is plotted as a function of $s$ at different time steps for (c) $N=256$, and (d) $N=512$. Here, one timestep $t$ constitutes $m=\mathrm{log}_{2}N$ even and $m=\mathrm{log}_{2}N$ odd circuit iterations. Black dotted vertical lines highlight the window over which $\mathcal{I}_3$ is plotted as a function of $s$ in (a) and (b). The numerical results are averaged over 3000 random circuit realizations.
\label{Fig:Different_Duration}
}
\end{figure}
As mentioned in the main text, increasing the time step $t$, increases the many-body entanglement generated in the system. This is reflected in the appreciable negativity of the tripartite mutual information. As $t$ increases, the transition between slow scrambling to fast scrambling regime is more evident only in larger system sizes as shown in \figref{Fig:Different_Duration} (a) and (b) for the neutral atoms-based model, as smaller systems sizes get entangled at a lower $t$. For a fixed system size, increasing the time step highlights a stark difference in $\mathcal{I}_3$ as a function of the tunable parameter $s$. With increasing time steps, $\mathcal{I}_3$ becomes larger and negative for $N=256$ and $N=512$ near $s=0$ as shown in \figref{Fig:Different_Duration} (c) and (d). For the numerical simulations to show transition in \figref{Fig:Different_Duration} (a) and (b), we stick to the interval highlighted by black vertical lines in \figref{Fig:Different_Duration} (c) and (d).

\section{Haar Random Limit}
\label{APP:Heuristic_toy}
In this section, we analyze the Clifford circuit models in the limit where two-qubit gates are replaced with 
Haar random gates. In this limit, the $0$\textsuperscript{th} order Renyi entropy, $S_{(0)}$ (the rank of the reduced density matrix),
can be computed by the minimum cut through the tensor network representation of the circuit \cite{Skinner2019}. 
In this section of the appendix, we use this mapping to show: 
first, this transition does not occur with the simple creation of boundary-crossing entangled qubit,
and second to give provide the analytical upper bound to the critical point $s_c$ of the Weighted random All-to-All (WrAA) model.

\subsection{Impossibility of achieving the negativity in tripartite mutual information from Bell pair creation}
We first investigate the scenario where every qubit in subregions $A$, $B$, $C$, and $D$ 
defined in the main text (\figref{fig:overview}(a))
forms a pair with a qubit outside the subregion (e.g.~a qubit in $A$ pairing with a qubit in subregion $B$, $C$ or $D$ 
but does not pair with other qubits in the subregion $A$). 
This is the scenario where the entanglement entropy of the given subregion is maximized, 
and hence the information in a subregion is completely delocalized just from the non-overlapping formation of bell pairs.
We show that even in this scenario, the tripartite mutual information can not be negative. 

In this scenario, we will have $S^{(0)}_X=N/4$
and $S^{0}_{XY}=N/2-2N_{XY}$, where $X\neq Y$ are any subregion and $N_{XY}$ is a bell pairs within the subregion.
In the case of the bell pairs, the tripartite mutual information evaluated with $n$\textsuperscript{th} order Renyi entropy is 
\begin{align}
   I^{(0)}_3= I^{(2)}_3=I^{(\mathrm{v.N.})}_3 = 2(N_{AB}+N_{AC}) - N/2 + 2N_{BC}.
\end{align}
Here $I^{0}$, $I^{2}$, and $I^{\mathrm{v.N.}}$ are tripartite mutual information calculated with 
Renyi-$0$, Renyi-$2$, and von Neumann entropy.
However, from the constraint, we have relations: $N_{AB}+N_{AC}=N/4-N_{AD}$ and $N_{BC}=N_{AD}$. Substitution of these yields:
\begin{align}
   I^{(0)}_3= I^{(2)}_3=I^{(\mathrm{v.N})}_3 = 0.
\end{align}

\subsection{Onset of scrambling as a dynamical transition characterized by counting discrete gates}
In the previous subsection, we show that the formations of the inter-regional 
Bell-pairs are not enough for the tripartite mutual information to become negative. 
Therefore, in this subsection, we expand our analysis to the three-qubit entanglement between three unique subregions in the system.

For the information to be delocalized between three subregions $A$, $B$, and $C$ in the output set of qubits, 
there should be gates that cross both the boundaries of a given region. 
Thus by calculating the probability $P^{*}(s)$ of at least two gates applied to a qubit in a given region that crosses both boundaries over multiple circuit realizations, 
we can comment on the extent of the proliferation of entanglement in the system and provide a lower bound to the value of $s_c$ of a given model. 
Non-existence of such gate configurations in the thermodynamic limit, $P^{*}(s) \rightarrow 0$, emphases that information is localized in a given region and not globally embedded across the system. 

Here we derive $P^{*}(s)$ analytically as a function of $s$ for WrAA circuit discussed in the main text. 
Before we compute $P^{*}_{\mathrm{WrAA}}(s)$ for WrAA, 
we first compute $P^{*}_{\mathrm{WrAA}}(i,s)$, the probability of two gates applied on $i$\textsuperscript{th} qubit from the boundary of region $A$ which entangles two discrete regions outside of $A$. 
This can be written as the following:
\begin{align}
   P^{*}_{\mathrm{WrAA}}(i,s)  = J^2_{s,\mathrm{WrAA}} \left( 
      \left( \sum_{k=i+1}^{N/2}k^{s} \right) \left( \sum_{l=N/4-i}^{N/2} l^s\right)
- \left( \frac{N}{2}  \right)^{2s} \right)
+ 
   J^2_{s,\mathrm{WrAA}} \left( 
      \left( \sum_{k=i+1}^{N/4}k^{s} \right) \left( \sum_{l=i+1+N/4}^{N/2} l^s\right)
- \left( \frac{N}{2}  \right)^{2s} \right)
\end{align}
Assuming that the largest contributions on $P^{*}(s)$ of the entire region are coming from the bulk qubits in region $A$, namely $1 \ll i\ll \frac{N}{4}$, 
The summations can be simplified using the Euler-Maclaurin formula \cite{Apostol1998,nistDLMFxA7252}, 
one obtains the following approximation for $s<-1$:
\begin{align}
   \sum_{k=1}^{N} k^s = \zeta(-s)  - \frac{N^{s+1}}{-s-1} + \mathcal{O}(N^{s}).
\end{align}
With this approximation, the function $P^{*}_{\mathrm{WrAA}}(i,s)$ can be approximated as 
\begin{align}
   P^{*}_{\mathrm{WrAA}}(i,s) &\approx \frac{J^2_{s,\mathrm{WrAA}}}{\left( s+1  \right)^2} 
   \left[ \left( \frac{N}{2}  \right)^{2s+2} 
   - \left( \frac{N}{2}  \right)^{s+1}\left( \left( i+1 \right)^{s+1} + \left( \frac{N}{4}-i  \right)^{s+1}   \right) 
   + \left(\left( i+1  \right)\left( \frac{N}{4}-i  \right)\right)^{s+1}    \right] \nonumber\\
                              &+ \frac{J^2_{s,\mathrm{WrAA}}}{\left( s+1  \right)^2} 
                              \left[ \left( \frac{N}{2\sqrt{2}}  \right)^{2s+2} 
                                 - \left( \frac{N}{4}  \right)^{s+1}\left( 2^{s+1} \left( i+1 \right)^{s+1} + \left( i+1+\frac{N}{4} \right)^{s+1}   \right) 
   + \left(\left( i+1  \right)\left( i+1+\frac{N}{4} \right)\right)^{s+1}    \right].
\end{align}
$P^{*}_{\mathrm{WrAA}}(s)$ that we want is then, the $P(i,s)$ summed over the $i$ in region $A$. In the thermodynamic limit, this becomes
\begin{align}
   \lim_{N\to\infty} P_{\mathrm{WrAA}}^*(s) = 
   \begin{cases}
      0 & (s<-1.5) \\
      \frac{5-9\sqrt{2}+2\pi+4\ln(1+\sqrt{2})}{2\zeta^2(1.5)} &(s=-1.5) \\
      1 & (-1.5<s)
   \end{cases},
   \label{Eq:Prob_cases}
\end{align}
indicating that proliferation of entanglement in the system does not occur in the system in the regime $s<-s_c$. 

\begin{figure}
    \centering
    \includegraphics[width=1.0\columnwidth]{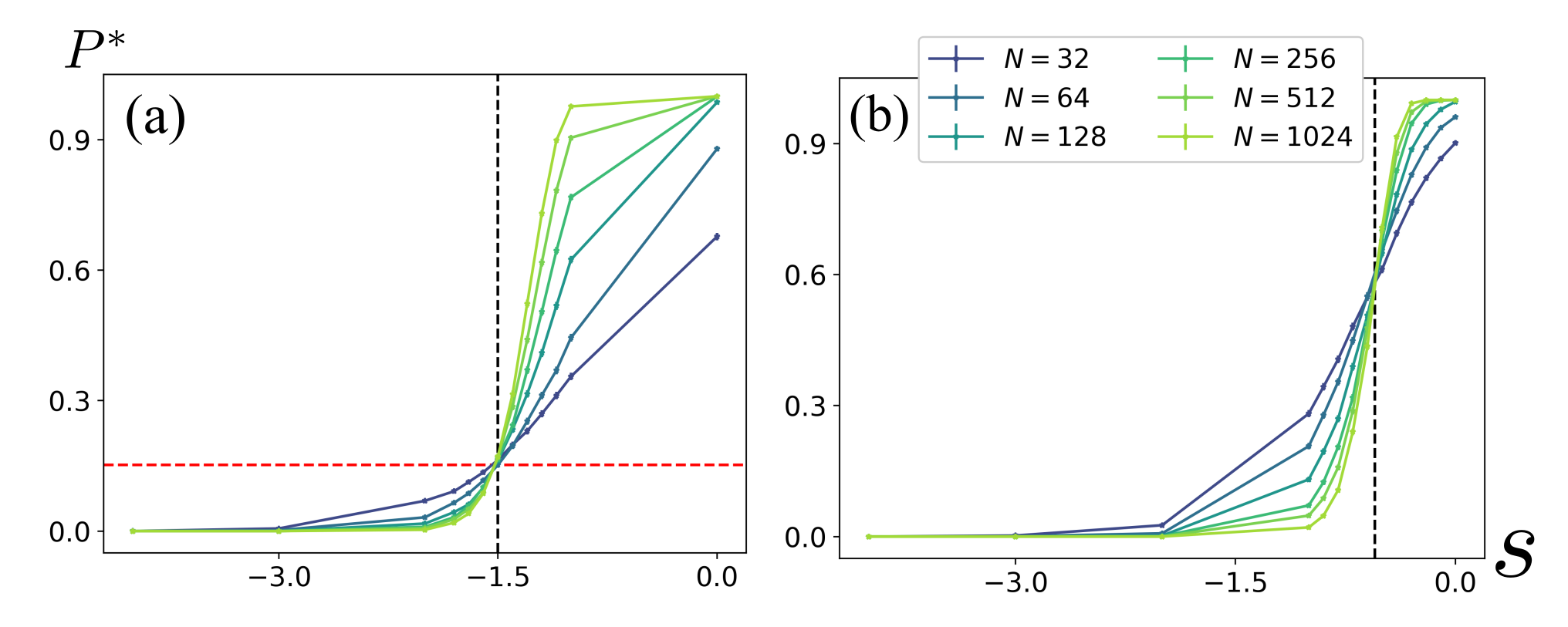}
    \caption{The probability $P^{*}$ of having at least two gates crossing both the boundaries of a given subregion as a function of a tunable parameter $s$ for system sizes $N=32,64,..1024$ for (a) weighted random all-to-all model (WrAA) and (b) neutral atoms-based model discussed in the main text. The critical point characterized by the tripartite mutual information for the WrAA model was $s_c=-1.33$. Here we see a transition around (a) $s_c=-1.5$ (black dashed line). The red dashed line is the theoretical value of $P_{\mathrm{WrAA}}^*(s)$ as given in \eqeqref{Eq:Prob_cases}. Similarly, the transition in $\mathcal{I}_3$ in the neutral atoms-based model had a critical point $s_c=-0.33$. However, the transition characterized by the probability $P^{*}$ in this case has a transition around (b) $s_c=-0.5$.}
    \label{fig:simpler_model}
\end{figure}

Finally, we plot, in \figref{fig:simpler_model}, the probability $P^{*}(s)$ 
for WrAA model and sparsely coupled neutral atom based model discussed in the main text. 
The value of the analytical transition point is indicated with a red dotted line in the figure.
The results are averaged over 3000 circuit realisations.

\section{Brownian Model}

\label{app:Brownian}

Here we analyze the dynamical $\mathcal{I}_3$ transition using a chain of coupled Brownian circuit clusters. The model comprises $N$ clusters, each consisting of a large number of spins $M$. We work in a mean-field limit with $M \rightarrow \infty$, such that each cluster behaves like a single classical degree of freedom with quantum fluctuations suppressed by $O(1/\sqrt{M}) \rightarrow 0$. The plan is to introduce two copies of the system to measure the second Renyi entropy, take the disorder average over the Brownian couplings, and express the result as an effective Hamiltonian $H_{\mathrm{eff}}$. We then consider the limit $\mathcal{J} t \gg 1 \gg b$, representing strong scrambling within each cluster and weak scrambling between clusters, and show that the model maps onto a long-range Ising model in this parameter regime. Finally, in the thermodynamic limit $N \rightarrow \infty$, the long-range Ising model exhibits a known mean-field ordering transition

Consider a system of $M \times N$ spin-1/2 qubits $\vec{S}_{iu} = S^{\alpha}_{iu}$ where $u,v = 0,\ldots,M-1$ label spins within each cluster and $i,j = 0,\ldots,N-1$ label the clusters, which are arranged in a 1D chain with closed boundary conditions. We drive this system with Brownian unitary dynamics at each timestep $dt$:
\begin{align}
    U_t &= e^{-i H(t) dt} \\
    &= \exp \left[- i J_{uv}^{\alpha \beta}(t) S^{\alpha}_{iu} S^{\beta}_{iv} dt - i K_{iu,jv}^{\alpha \beta}(t) S^{\alpha}_{iu} S^{\beta}_{jv} dt \right], \nonumber
\end{align} 
where the repeated indices are implicitly summed over and $J,K$ are white-noise couplings that generate intra- and inter-cluster interactions, respectively. These couplings have zero mean and variance
\begin{align}
    \mathbb{E} \left[ J_{uv}^{\alpha \beta}(t) J_{u'v'}^{\alpha' \beta'}(t') \right] &= \frac{\mathcal{J}}{M dt} \left(1 - b \right) \delta^{\alpha \alpha'} \delta^{\beta \beta'} \delta_{uu'} \delta_{vv'} \delta_{t t'} \\ \nonumber
    \mathbb{E} \left[ K_{iu,jv}^{\alpha \beta}(t) K_{i'u',j'v'}^{\alpha' \beta'}(t') \right] 
    &= \frac{\mathcal{J}}{M dt} b A_{ij} \ \delta^{\alpha \alpha'} \delta^{\beta \beta'}
    \delta_{ii'} \delta_{jj'} \delta_{uu'} \delta_{vv'} \delta_{t t'} \nonumber        
\end{align}
where $\mathcal{J}$ controls the overall coupling strength, $A_{ij}$ is the (normalized) adjacency matrix of the inter-cluster couplings, and $0 \leq b \leq 1$ controls the ratio of the intra- and inter-cluster couplings. The factor of $M$ is necessary to ensure that the instantaneous Hamiltonian $H(t)$ is extensive. The full evolution over a total time $t$ is given by the unitary matrix $U = \prod_{0}^t U_t$, and we work in the limit as $dt \rightarrow 0$ with $\mathcal{J} t$ fixed. We will also take the limit $M \rightarrow \infty$, so the model is completely controlled by the coupling matrix $A_{ij}$, the chain length $N$, and the parameter $b$.

To calculate $\mathcal{I}_3$, we must compute the second Renyi entropy of a subregion $A$. This is achieved by introducing two copies of the system and performing a SWAP test on the qubits in region $A$, which is equivalent to computing the expectation value:
\begin{equation}
    \label{eq:secondrenyi}
    S^{(2)}_A = - \ln \mathrm{Tr} \left[ \left(\mathrm{SWAP}_{A_1 A_2} \right)  U \rho_0 \adj{U} \otimes U \rho_0 \adj{U} \right]
\end{equation}
where $\rho_0$ is the initial state. Using the Choi-Jamiolkowski isomorphism (i.e.~`flip the bras to kets'), we map the mixed-state dynamics of two copies onto the pure-state dynamics of four copies:
\begin{equation}
    S^{(2)}_A = - \ln \llangle \mathrm{SWAP}_{A_1 A_2} \vert U \otimes U^{\tau} \otimes U \otimes U^{\tau} \lvert \rho_0 \rrangle
\end{equation}
where
\begin{equation}
    U^{\tau} = \sigma_y U^* \sigma_y
\end{equation}
is the time-reversal of $U$ (the additional factors of $\sigma_y$ are necessary so that spins transform correctly $\vec{S} \rightarrow - \vec{S}$ under time-reversal). In particular, $(U_t)^{\tau} = e^{+i H(t) dt}$ because the Hamiltonian is quadratic in the spins. In this way, the Renyi entropy can be understood as a bulk propagator $U \otimes U^{\tau} \otimes U \otimes U^{\tau}$ connecting the initial state $\lvert \rho_0 \rrangle$ to the final state $\lvert \mathrm{SWAP}_{A_1 A_2} \rrangle$. We label the four copies (or `replicas') of the system by the indices $r,s = 1,2,3,4$.

\subsection{Disorder Average}

We focus first on the dynamics of the bulk propagator, and later consider the effects of the initial and final states. Because the Brownian coefficients are uncorrelated in time, we can perform the disorder average independently at each timestep:
\begin{align}
    \mathbb{E} \left[ U_t \otimes U_t^{\tau} \otimes U_t \otimes U_t^{\tau} \right] \approx 1 &- \frac{\mathcal{J}}{M dt}(1-b) \sum_{r < s} (-1)^{r+s} \sum_{i} \left( \sum_{u} \vec{S}_{iu} \cdot \vec{S}_{iu} \right)^2 dt^2 \\ \nonumber
    &- \frac{\mathcal{J}}{M dt} b \sum_{r < s} (-1)^{r+s} \sum_{ij} A_{ij} \left( \sum_u \vec{S}_{iu} \cdot \vec{S}_{iu} \right) \left( \sum_v \vec{S}_{jv} \cdot \vec{S}_{jv} \right) dt^2
\end{align}
where we have expanded to second order in the small quantity $dt$. We stack these timesteps together to find that the bulk propagator is governed by a quantum statistical mechanics model with Gibbs weight $\exp(-t H_{\mathrm{eff}})$ and effective Hamiltonian
\begin{equation}
    \label{eq:Heff}
    H_{\mathrm{eff}} = M \mathcal{J} \sum_{r<s} (-1)^{r+s} \left[ (1-b) \sum_{i} \left( G_i^{rs} \right)^2
    + b \sum_{ij} A_{ij} G_i^{rs} G_j^{rs} \right]
\end{equation}
where 
\begin{equation}
    G_i^{rs} := \frac{1}{M} \sum_{u} \vec{S}^r_{iu} \cdot \vec{S}^s_{iu}
\end{equation}
are the mean fields on each cluster, and where the total time $t$ plays the role of inverse temperature (so larger time $t$ means colder temperature).

Note that this derivation implicitly assumes that we can take the expectation value inside the logarithm, i.e. $\mathbb{E} \ln x \approx \ln \mathbb{E} x$. We generally expect this to be true at large $M \rightarrow \infty$ \cite{bentsen2021measurement}, but this must be checked explicitly by calculating higher moments and verifying that fluctuations are negligible. An exact expression for $\mathbb{E} \ln x$ can be derived in principle using a replica limit.

To make further progress, it is convenient to switch to a path integral representation
\begin{equation}
    \llangle \psi_T \rvert \exp(-t H_{\mathrm{eff}}) \lvert \psi_0 \rrangle = \int \mathcal{D} F_i^{rs} \mathcal{D} G_i^{rs} \mathcal{D} S^{r\alpha}_{iu} \exp(-M \mathcal{S})
\end{equation}
with action
\begin{align}
    \mathcal{S} = \int_0^t dt &\left[ \mathcal{J} (1-b) \sum_{r < s} (-1)^{r+s} \sum_{i} \left( G_i^{rs} \right)^2
    + \mathcal{J} b \sum_{r < s} (-1)^{r+s} \sum_{ij} A_{ij} G_i^{rs} G_j^{rs} \right. \\ \nonumber
    &\left. - \sum_{r<s} \sum_i iF_{i}^{rs} \left( G_i^{rs} - \frac{1}{M} \sum_{u} \vec{S}^r_{iu} \cdot \vec{S}^s_{iu} \right) \right]
\end{align}
where we have introduced Lagrange multipliers $i F^{rs}_i$ to enforce the mean-field constraint. Here the mean fields $G_i^{rs}$, Lagrange multipliers $F_i^{rs}$, and spin variables $S_{iu}^{\alpha,r}$ are all functions of Euclidean time $t$, and the spin variables satisfy the boundary conditions imposed by $\lvert \psi_{0,t} \rrangle$.
Importantly, we have buried the kinetic term for the  spin variables in the integration factor $\int \mathcal{D} S^{\alpha}$. Now that the spins have been isolated to a single term it is convenient to separate them from the rest of the action, so we write
\begin{equation}
    \mathcal{S} = \int_0^t dt \left[ \mathcal{J} \sum_{r < s} (-1)^{r+s} \sum_{ij} \chi_{ij} G_i^{rs} G_j^{rs} - \sum_{r<s} \sum_i iF_{i}^{rs} G_i^{rs} \right] - \sum_i \ln \mathcal{K}_i
\end{equation}
where
\begin{align}
    \mathcal{K}_i := &\int \mathcal{D} S^{\alpha,r} \exp \left(-  \int_0^t dt \sum_{r<s} i F_i^{rs} \vec{S}^r \cdot \vec{S}^s \right) \nonumber \\
    = & \llangle \psi_t \rvert \prod_t \exp \left(- \sum_{r<s} iF_i^{rs}(t) \vec{S}^r \cdot \vec{S}^s \right) \lvert \psi_0 \rrangle
\end{align}
is the propagator for a single spin cluster under the mean field Lagrange multiplier $iF_i^{rs}$, and
\begin{equation}
    \chi_{ij} := (1-b)\delta_{ij} + b A_{ij}
\end{equation}
where we have assumed that $A_{ij}$ is translation invariant, symmetric, and normalized.
By setting $b = 0$ we recover $N$ identical decoupled copies of the original single-cluster model.

\subsection{Mapping to Long-Range Ising Model}

We now consider the limit $\mathcal{J} t \gg 1 \gg b$ and show that the dynamics of the effective Hamiltonian in this regime is equivalent to a long-range Ising model at finite inverse temperature $t$. The propagator $\mathcal{K}_i$ is simply the quantum mechanics of $2k = 4$ spins $r,s = 1,2,3,4$ in the external fields $F^{rs}_i(t)$. The SWAP boundary condition at the final time $t$ projects the system onto the spin singlet subspace $S_{\mathrm{Tot}}^2 = 0$, where $\vec{S}_{\mathrm{Tot}} = \sum_r \vec{S}^r$. Moreover, since the Heisenberg couplings $\vec{S}^r \cdot \vec{S}^s$ manifestly conserve the total spin, we see that the entire dynamics is restricted to the singlet $S_{\mathrm{Tot}}^2 = 0$ subspace. For $k = 2$ this subspace is spanned by the basis vectors
\begin{align}\label{eq:basis2spins}
    \ket{\uparrow} &= \frac{1}{2 \sqrt{3}} \left(2\ket{1010} + 2\ket{0101} - \ket{0011} - \ket{1100} - \ket{1001} - \ket{0110} \right) \nonumber \\
    \ket{\downarrow} &= \frac{1}{2} \left( \ket{0011} + \ket{1100} - \ket{1001} - \ket{0110} \right)
\end{align}

Thus, each propagator can be written as the dynamics of a spin-1/2 particle in a time-dependent magnetic field:
\begin{equation}
    \mathcal{K}_i = \bra{\psi_T} \exp{\left[ \frac{1}{2} \int_0^t dt \ \vec{B}_i(t) \cdot \vec{\sigma} \right]} \ket{\psi_0} e^{B^0_i t / 2}
\end{equation}
where the magnetic field has components
\begin{align}
    B^x_i &= \frac{\sqrt{3}}{2} \left(i F^{12}_i + i F^{34}_i - i F^{14}_i - i F^{23}_i \right) \nonumber \\
    B^y_i &= 0 \nonumber \\
    B^z_i &= \frac{1}{2} \sum_{r<s} i F^{rs}_i - \frac{3}{2} \left( i F^{13}_i + i F^{24}_i \right) \nonumber \\
    B^0_i &= \frac{1}{2} \sum_{r<s} iF_i^{rs}
\end{align}
At sufficiently long times $\magn{\vec{B}}t \gg 1$, the boundary conditions do not affect the bulk physics, and the bulk is determined by its time-independent saddle points $\vec{B}_i(t) = \vec{B}_i$. In this limit, only the ground state contributes, and the propagator, therefore, simplifies to:
\begin{equation}
    \mathcal{K}_i \approx c e^{\magn{\vec{B}_i} t /2} e^{B_i^0 t/2}
\end{equation}
where the unimportant constant factor $c$ comes from the overlap of the initial state with the spin singlet subspace basis vectors $\ket{\uparrow},\ket{\downarrow}$.

\begin{figure}
    \centering
    \includegraphics[width=0.3\textwidth]{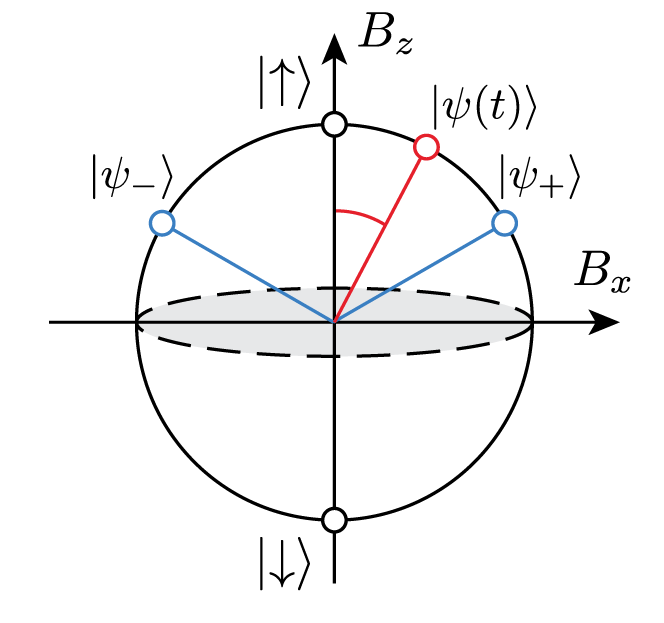}
    \caption{The effective spin-1/2 particle governing the propagator $\mathcal{K}_i$ can be visualized as a time-dependent state $\ket{\psi(t)}$ (red) on a Bloch sphere spanned by the basis vectors $\ket{\uparrow}, \ket{\downarrow}$. The saddle-point solutions (blue) are the vectors $\psi_{\pm} = \frac{\sqrt{3}}{2} \ket{\uparrow} \pm \frac{1}{2} \ket{\downarrow}$}
    \label{fig:replicaspinblochsphere}
\end{figure}

Therefore, our action simplifies to:
\begin{equation}
    \mathcal{S} = \mathcal{J}t \sum_{r < s} (-1)^{r+s} \sum_{ij} \chi_{ij} G_i^{rs} G_j^{rs} - \sum_{r<s} \sum_i iF_{i}^{rs} G_i^{rs} t - \sum_i \left( \magn{\vec{B}_i} + B_i^0 \right) t / 2
\end{equation}
The final step is to find the saddle-points of this action by taking derivatives $\partial \mathcal{S} / \partial G = \partial \mathcal{S} / \partial F = 0$ with respect to the fields $F,G$. The saddle-point equations of motion are
\begin{align}
    \label{eq:eom}
    iF_i^{rs} &= 2 \mathcal{J} (-1)^{r+s} \sum_j \chi_{ij} G_j^{rs} \nonumber \\
    G_i^{rs} &= -\frac{1}{2} \frac{\partial}{\partial i F_i^{rs}} \left( \magn{\vec{B}_i} + B_i^0 \right)
\end{align}
We may immediately eliminate the fields $G_i^{rs}$ from these equations. We also notice that the antisymmetric combinations $iF_i^{12}-iF_i^{34},iF_i^{14}-iF_i^{23},iF_i^{13}-iF_i^{24} = 0$ vanish identically, so we introduce the symmetric combinations
\begin{align}
    \ph_i &= i F_i^{12} + i F_i^{34} \\
    \pv_i &= i F_i^{14} + i F_i^{23} \nonumber \\
    \px_i &= i F_i^{13} + i F_i^{24}  \nonumber
\end{align}
which simplifies the equations of motion to
\begin{align}
    \label{eq:eoms}
    \ph_i &= 2\mathcal{J} \sum_j \chi_{ij} \left(\frac{\partial{\magn{\vec{B}_j}}}{\partial \ph_j} + \frac{1}{2} \right) = \mathcal{J} \sum_j \chi_{ij} \left(\frac{1}{\magn{\vec{B}_j}} \left( 2 \ph_j - \pv_j - \px_j \right) + 1 \right) \\
    \pv_i &= 2\mathcal{J} \sum_j \chi_{ij} \left(\frac{\partial{\magn{\vec{B}_j}}}{\partial \pv_j} + \frac{1}{2} \right) = \mathcal{J} \sum_j \chi_{ij} \left(\frac{1}{\magn{\vec{B}_j}} \left( - \ph_j + 2\pv_j - \px_j \right) + 1 \right) \nonumber \\
    \px_i &= -2\mathcal{J} \sum_j \chi_{ij} \left(\frac{\partial{\magn{\vec{B}_j}}}{\partial \px_j} + \frac{1}{2} \right) = -\mathcal{J} \sum_j \chi_{ij} \left(\frac{1}{\magn{\vec{B}_j}} \left( - \ph_j - \pv_j + 2\px_j \right) + 1 \right) \nonumber
\end{align}
We may solve these equations order-by-order in the small parameter $b$. At the lowest order in $b$, we subtract the first two equations to find that $\magn{\vec{B}_i} = 3 \mathcal{J} + O(b)$. We also find the linear combination
\begin{equation}
    \ph_i + \pv_i - \px_i = 3\mathcal{J} \sum_{j} \chi_{ij} = 3 \mathcal{J} - 3 b \mathcal{J} \sum_j \left(\delta_{ij} - A_{ij} \right) 
\end{equation}
Substituting this into the last line of \eqref{eq:eoms}, we conclude that $\phi_i^c$ vanishes at lowest order $\phi_i^c = 0 + O(b)$. Finally, we substitute $\ph_i + \pv_i = 3 \mathcal{J} + O(b)$ into $\magn{\vec{B}_i} = 3 \mathcal{J} + O(b)$ to solve for
\begin{equation}
    \ph_i - \pv_i = \pm 3 \mathcal{J} + O(b)
\end{equation}

Thus, we find exactly two saddle points $\ket{\psi_{\pm}}$ for each cluster $i$, which we label with an Ising spin $\sigma_i = \pm 1$. In the spin-1/2 replica subspace these states are written
\begin{equation}
    \ket{\psi_{\pm}} = \frac{\sqrt{3}}{2} \ket{\uparrow} \pm \frac{1}{2} \ket{\downarrow}
\end{equation}
When $\sigma_i = +1$ we find $G_i^{12} = G_i^{34} = -3/4 + O(b)$ with all other fields vanishing, while for $\sigma_i = -1$ we find $G_i^{13} = G_i^{24} = -3/4 + O(b)$ with all other fields vanishing. Plugging these saddle-point solutions back into the action, we find that the bulk dynamics is
\begin{align}
    \exp(-t H_{\mathrm{eff}}) &\approx \sum_{\{\sigma_i = \pm1 \}} e^{-t H[\sigma]} \nonumber \\
    H[\sigma] &= \mathcal{J} \sum_{ij} \chi_{ij} (3/4)^2 \left( \sigma_i \sigma_j + 1 \right)
\end{align}
Apart from additive constants, this is simply a long-range Ising model with coupling matrix $\chi_{ij}$. To get the true action, in principle we ought to expand to higher powers of $b$, but doing so will not change the number of saddle-point solutions, it will only modify the energy of the effective Ising model.

So far we have discussed the bulk dynamics of $H_{\mathrm{eff}}$ while ignoring the boundary conditions $\lvert \psi_{0,t} \rrangle$. The effect of these boundary conditions is to impose an energy penalty coming from the overlap of the bulk saddle points $\ket{\psi_{\pm}}$ with the boundary states. We may treat each cluster $i$ separately. For the initial condition $\lvert \psi_{0} \rrangle =  \ket{\uparrow}$, we have an overlap
\begin{equation}
    \bra{\psi_{\pm}} \psi_0 \rrangle = \frac{\sqrt{3}}{2}
\end{equation}
independent of the choice of saddle point. Because this overlap doesn't depend on the saddle point $\sigma_i = \pm1$, this gives an unimportant overall constant shift to the action which we ignore. By contrast, the final condition $\lvert \psi_{t} \rrangle$ depends on whether the cluster belongs to the SWAP-ed region $A$ or the un-SWAP-ed region $\overline{A}$. If we pick a cluster $i \in A$, then the overlap is 
\begin{align}
    \llangle \psi_t \ket{\psi_{+}} &= \langle{\psi_-}\ket{\psi_+} = \frac{1}{2} \nonumber \\
    \llangle \psi_t \ket{\psi_{-}} &= \langle{\psi_-}\ket{\psi_-} = 1
\end{align}
and vice-versa when $i \in \overline{A}$. Thus, whenever there is a mismatch between the bulk saddle point and the final boundary condition, we have an energy penalty $-\log 2$ in the effective action. This is equivalent to having an applied magnetic field $h_i$ acting on each spin $\sigma_i$ in the long-range Ising model. Hence we may write
\begin{equation}
    \llangle \psi_T \rvert \exp(-t H_{\mathrm{eff}}) \lvert \psi_0 \rrangle \approx \sum_{\{\sigma_i = \pm1 \}} e^{-t H[\sigma] - \log 2 \sum_{i} ( 1- h_i \sigma_i)/2}
\end{equation}
where $h_i = -1$ for $i \in A$ and $h_i = +1$ for $i \in \overline{A}$.

To summarize, we started with a Brownian model defined by parameters $N,M,\mathcal{J}t,b,A_{ij}$, and we considered the limit $M \rightarrow \infty$, followed by $\mathcal{J} t \gg 1 \gg b$ such that $b \mathcal{J} t \sim O(1)$. Finally, we consider the thermodynamic properties of the resulting classical stat-mech model in the thermodynamic limit $N \rightarrow \infty$.

\end{document}